%% file: main.tex
\documentclass{article}

\pdfoutput=1
\usepackage[utf8]{inputenc}
\usepackage[T1]{fontenc}
\usepackage{lmodern}
\usepackage[a4paper]{geometry}

\usepackage[sorting=none]{biblatex}
\usepackage{blindtext}
\usepackage{tikz}
\usepackage{pgfplots}
\pgfplotsset{compat=1.18}
\usepackage{tcolorbox}

\usepackage{mystyle}  
\usepackage{mymacros}  

\title{Integrability in Gravity from Chern-Simons Theory}
\author{Lewis T.\ Cole and Peter Weck}

\begin{document}

\hypersetup{linkcolor=mytextcolour}
\input{titlepage}
\setcounter{page}{2}
\tableofcontents
\hypersetup{linkcolor=mylinkcolour}

\newpage

\input{text}

\newpage

\appendix

\input{appendix}

\newpage

\printbibliography

\end{document}

%% file: titlepage.tex
\begin{titlepage}

\vspace*{2cm}

\begin{center}
    {\Large \textbf{Integrability in Gravity from Chern-Simons Theory}}

    \vspace{1cm}

    {\large \phantom{$^{a,1}$} Lewis T.\ Cole\footnote{\texttt{l.t.cole@pm.me}} and Peter Weck\footnote{\texttt{p.j.weck@swansea.ac.uk}}}

    \vspace{0.5cm}

    {\textit{Department of Physics, Swansea University, \\ Swansea SA2 8PP, United Kingdom \\}}
\end{center}

\vspace{1cm}

\begin{abstract}
    This paper presents a new perspective on integrability in theories of gravity. 
    We show how the stationary, axisymmetric sector of General Relativity can be described by the boundary dynamics of a four-dimensional Chern-Simons theory. 
    This approach shows promise for simplifying solution generating methods in both General Relativity and higher-dimensional supergravity theories. 
    The four-dimensional Chern-Simons theory presented generalises those for flat space integrable models by introducing a space-time dependent branch cut in the spectral plane.
    We also make contact with twistor space approaches to integrability, showing how 
    the branch cut defects of four-dimensional Chern-Simons theory arise from a discrete reduction of six-dimensional Chern-Simons theory. 
\end{abstract}

\end{titlepage}

%% file: text.tex
\section{Introduction}

Many advances in our understanding of theories of gravity have been closely linked to the study of integrability. 
This is particularly true in the context of black hole solutions. The integrable structure of gravity with $D-2$ commuting Killing symmetries facilitates powerful solution generating methods, which have led to the discovery of a surprisingly rich array of exact solutions \cite{Emparan:2008eg}. Many black hole uniqueness theorems also rely on this integrable structure \cite{Mazur:1982db,Hollands:2007aj, Hollands:2008fm, Lucietti:2020ltw, Lucietti:2020phh}. 

Two formal ingredients underlie many of these developments.
First, with this much spacetime symmetry, the equations of various $D$-dimensional theories of gravity reduce to those of nonlinear $\sigma$-models in two dimensions \cite{Breitenlohner:1986um,Breitenlohner:1987dg}. The fundamental field of these 2d models roughly corresponds to the internal part of the metric and takes values in a coset group. In the canonical example, four-dimensional General Relativity (GR) is reduced by axial $\partial_\phi$ and stationary $\partial_t$ Killing vectors to a $\sigma$-model with target space $\mathrm{SL}(2,\mathbb{R})/\mathrm{SO}(2)$, coupled to 2d gravity. Unlike the usual symmetric-space $\sigma$-model, it features a spacetime-dependent coupling parameter which breaks translational invariance. 
The second ingredient is that these 2d models are integrable, a fact which is exhibited by the existence of a flat `Lax' connection expressed in terms of an auxiliary complex parameter. 

The goal of this paper is to show how recent advances relating 2d integrable models to 4d Chern-Simons theories \cite{Costello:2019tri, Delduc:2019whp} can be applied to stationary and axisymmetric GR (see \cite{Lacroix:2021iit} for a pedagogical introduction to 4dCS theory). This framework beautifully captures the integrable structure of gravity in the language of gauge theory and geometry. The Lax connection is realized as a dynamical gauge field, and the `spectral' parameter as a spatial coordinate. 
We will demonstrate how the 2d $\sigma$-model for stationary, axisymmetric gravity emerges from the boundary dynamics of this Chern-Simons theory. 
We believe this framework will provide fertile ground for the study of powerful but highly technical solution generating methods, not only in GR but also in higher-dimensional supergravity theories. Below we provide two divergent but related motivations for this work: one which may appeal to researchers interested in solution generating methods in theories of gravity, and the other to those interested in applications of 4d and 6d Chern-Simons theory to lower-dimensional integrable models. 

\subsection{Motivation from exact techniques in gravity}  
Let us first consider the application of the Lax formalism to solution generating methods in theories of gravity.
As early as 1978, the Lax formalism was used by Belinski and Zakharov to generate 4d vacuum solutions describing $n$ black holes with a common axis of symmetry \cite{Belinsky:1971nt, Belinsky:1979mh}. The BZ method takes an initial solution $G_0$ for the metric on the Killing directions, defines an extension $G$ which depends on the spectral parameter $Z$, and constructs a new solution via a dressing transformation, 
 \begin{equation}
       G \mapsto \chi \, G, \qquad \chi =\boldsymbol{1}+\sum_{i=1}^{2n} \frac{\chi_i}{Z-Z_i}.
\end{equation}
Compatibility with the Lax strongly constrains the dressing matrix $\chi$. Its $Z$-dependence is shown above explicitly, and the form of $\chi_i$ and $Z_i$ is completely fixed\footnote{A few additional constraints are required beyond preservation of the form of the Lax. These ensure the resulting metric is symmetric, real-valued, and asymptotically well-behaved.} up to free parameters. These parameters set properties of the new solution such as mass, rotation, and NUT charge. The same method can be used when all the Killing vectors are spacelike, to generate various `multi-soliton' gravitational wave solutions.

In the application to 4d stationary and axisymmetric solutions, only the $n=1$ case of the Kerr black hole is free from non-physical features such as conical defects and closed time-like curves. 
However, the generalization of the BZ method (also known as the inverse scattering method) to higher dimensional GR \cite{Pomeransky:2005sj} has led to the discovery of a rich array of new solutions. Perhaps most fascinating are the black ring and black Saturn solutions, which have established that black holes of non-spherical horizon topology are a generic feature of vacuum gravity for $D>4$. See \cite{Emparan:2008eg} for a review of these developments. Extensions to Einstein-Maxwell theory have also been made in \cite{Alekseev:1980ew,Neugebauer:1983dp}. 

\begin{table} \label{coset_table}
      \centering
      \def\arraystretch{1.5}
      \begin{tabular}{l|l}
        \hline
       $D \geq 4$ vacuum gravity $\quad$ & $\quad \mathrm{SL}(D-2,\mathbb{R})/\mathrm{SO}(D-2)$ \\
       4d EMd gravity & $\quad \mathrm{SU}(2,1)/(\mathrm{SU}(2)\times \mathrm{U}(1))$ \\
        4d $N=4$ from 10d SuGra  $\quad$ & $\quad \mathrm{SO}(8,8)/(\mathrm{SO}(8)\times \mathrm{SO}(8))$\\
       4d $N=8$ from 11d SuGra  $\quad$ & $\quad \mathrm{E}_{8(+8)}/\mathrm{SO}(16)$\\
        \hline
      \end{tabular}
      \caption{Examples of $\sigma$-models obtained from dimensional reduction of different theories of gravity with $D-2$ commuting Killing vectors, from \cite{Breitenlohner:1987dg}. Note these integrable 2d models capture only the bosonic sector.}
\end{table}
    
These techniques have seen more limited application in supergravity theories even though the integrable 2d $\sigma$-model foundation widely applies. The BZ method is usually formulated directly in terms of components of the metric on Killing directions. While this is closely related to a field configuration in the coset space of the corresponding $\sigma$-model, direct application of the dressing transformation is not guaranteed to respect the coset group structure \cite{Figueras:2009mc}. Reconciling the BZ method and the $\sigma$-model approach pioneered by Breitenlohner and Maison can present a technical challenge --- see the minimal 5d supergravity case in \cite{Figueras:2009mc}, and STU supergravity models in \cite{Katsimpouri:2013wka}. Additionally, the dressing matrix for vacuum gravity with $D-2$ commuting Killing vectors is allowed only simple poles with residues of rank one. 
Higher rank residues are possible for more general coset groups \cite{Katsimpouri:2013wka}, further complicating the BZ methodology. 
One motivation to find simpler and more systematic ways to implement these transformations is to understand the scope of exact solutions in higher-dimensional (super)gravity theories. 

Another motivation comes from the black hole microstate problem, particularly the fuzzball proposal \cite{Bena:2022ldq, Bena:2013dka}. Some black hole microstates are coherent enough to admit a semi-classical description in terms of smooth and horizonless geometries. Most of the known examples of such `microstate geometries' are supersymmetric or represent small deformations away from supersymmetric solutions \cite{Bena:2006kb, Bena:2007kg, Bena:2015bea}. Methods to systematically generate more general stationary and axisymmetric microstate geometries would be quite significant. Recent progress on the static case has indirectly made use of inverse scattering and integral equation methods, employing 4d Einstein-Maxwell solutions to solve higher-dimensional equations of motion. This strategy was employed in \cite{Bah:2022yji} and \cite{Bah:2023ows} to construct the first examples of smooth and horizonless geometries which asymptotically resemble Schwarzschild black holes.

Accessing the full power of inverse scattering methods in supergravity theories may require approaches which are more systematic and better adapted to the $\sigma$-model presentation.
A promising new perspective on the Lax formalism and 2d integrable field theories (IFTs) has recently emerged, relating them to Chern-Simons (CS) theory on a 4d space with defects \cite{Costello:2019tri}. The Lax connection is treated as a dynamical gauge field in this extended space with the spectral parameter serving as a coordinate. 
Dressing transformations, such as in the BZ method, have the form of (singular) gauge transformations on the Lax connection, 
\begin{equation}
    \mathcal{L} \mapsto \chi \mathcal{L} \chi^{-1} - \dr \chi \chi^{-1} \ .
\end{equation}
These transformations may have a natural home in the 4dCS framework. In this paper, we lay the groundwork for further investigation along these lines by showing explicitly how the 2dIFT for stationary axisymmetric gravity can be derived from a novel 4dCS theory. While we often refer to the $\mathrm{SL}(2,\mathbb{R})/\mathrm{SO}(2)$ $\sigma$-model of the 4d vacuum case for the sake of concreteness, our construction is group agnostic. For example it applies equally to any of the coset group $\sigma$-models identified by Breitenlohner, Maison, and Gibbons in \cite{Breitenlohner:1987dg} (see e.g.\ table \ref{coset_table}).

\subsection{Motivation from 4d \& 6d Chern-Simons theory}

For the reader who is primarily interested in 4dCS and/or 6dCS theory, this paper presents several new results in these domains. 
The symmetric space $\sigma$-models describing integrable sectors of gravity have the unusual feature of a spacetime-dependent coupling (breaking translational invariance) and an associated Lax connection which mixes the spectral parameter with the spacetime coordinates. Accommodating these 2dIFTs into the 4dCS framework requires that we introduce several new ingredients. 
To highlight these novel features of our construction, we first recall that the action of 4dCS theory \cite{Costello:2019tri} is given by 
\begin{equation}
    S_{\text{4dCS}} = \frac{1}{2 \pi \rmi} \int_{\bCP^{1} \times \bR^{2}} \omega \wedge \tr \bigg( A \wedge \dr A + \frac{2}{3} A \wedge A \wedge A \bigg) \ , 
\end{equation}
where $\omega$ is a meromorphic 1-form. This theory allows for the systematic construction of a wide array of 2d integrable field theories as well as their associated Lax connections. Provided with the input data defining a 4dCS setup, namely a choice of 1-form $\omega$ and boundary conditions on the gauge field $A$, it is possible to localise the action to an integral over $\bR^{2}$. Different choices of input data will lead to different 2dIFTs, meaning that 4dCS theory is also a mechanism for exploring the space of integrable theories. For this reason, it is important to understand which properties of the input data are essential, and to find any extraneous constraints which may be relaxed. For example, it is not yet understood what the most general 1-form $\omega$ is which is both compatible with the localisation procedure and leads to an integrable field theory.

The existing literature on 4dCS theory has treated 1-forms which can be written in terms of a twist function as $\omega = \varphi(Z) \, \dr Z$, where $Z$ is a holomorphic coordinate on a complex curve, often taken to be $\bCP^{1}$. These 1-forms only have legs along $\dr Z$, and the twist function $\varphi(Z)$ is a meromorphic function of $Z$. In this paper, we will relax both of these assumptions, allowing $\omega$ to have legs along the spacetime directions, and the coefficients to have dependence on the spacetime coordinates, which we will often denote $\{ \rho, z \}$. The fact that our 4dCS theory retains its key properties after this generalisation is a surprising result. We will demonstrate that it is still possible to localise the action to a 2dIFT, and derive the associated Lax connection for the concrete example of stationary, axisymmetric GR.

Another perspective on this unfamiliar feature of our 4dCS theory is found by considering an alternative spectral parameter defined by 
\begin{equation}
    W = z + \frac{\rho}{2} \big( Z^{-1} - Z \big) \ . 
\end{equation}
The original $Z$-plane is a double covering of the $W$-plane and the meromorphic 1-form of interest may be expressed as $\omega = \dr W$. This presentation brings the 1-form into a more familiar form, but now the novelty lies in the relationship between $Z$ and $W$. In particular, the $W$-plane contains a branch cut between the points $W = z + \rmi \rho$ and $W = z - \rmi \rho$. From this perspective, we are generalising the \textit{branch cut defects} of \cite{Costello:2019tri} to allow for spacetime-dependent endpoints.

These generalisations of 4dCS theory were not the result inspired guesses, but were derived from a reduction of 6dCS theory. The 6-manifold underlying 6dCS theory \cite{Costello:2020, Costello:2021} is twistor space $\bPT$, which is isomorphic to $\bCP^{1} \times \bR^{4}$ as a real manifold. The action is given by 
\begin{equation}
    S_{\text{6dCS}} = \frac{1}{2 \pi \rmi} \int_{\bPT} \Omega \wedge \tr \bigg( \cA \wedge \bar{\pd} \cA + \frac{2}{3} \cA \wedge \cA \wedge \cA \bigg) \ , 
\end{equation}
where $\Omega$ is a meromorphic $(3,0)$-form. Much like the relationship between 4dCS theory and 2dIFTs, this six-dimensional counterpart localises to spacetime (i.e.\ $\bR^{4}$) producing four-dimensional integrable field theories (4dIFTs). The analog of the Lax connection for 4dIFTs is a self-dual Yang-Mills (sdYM) connection which may be derived from the 6dCS description.

Integrable models in lower dimensions are known to arise as conformal reductions of the 4d sdYM equations (see the textbook \cite{Mason:1996}). If we restrict to solutions which are invariant under two conformal Killing vectors, the remaining sdYM equations describe 2dIFTs. For example, it was noticed in \cite{Witten:1979tv} that the static, axisymmetric gravity equations are equivalent to a static, axisymmetric reduction of the sdYM equations. This observation inspired further work \cite{Ward:1982bf} in which the machinery of twistor theory was used to generate solutions to the vacuum Einstein's equations. These developments are discussed in the textbooks \cite{Fletcher:1990} and \cite{Mason:1996}.

When compared with this older literature on the topic, the advent of 6dCS theory gives us the opportunity to work at the level of an action, rather than just working with the equations of motion. This was recently applied in \cite{Penna:2020uky} to derive actions for 2dIFTs coming from dimensional reductions of gravity and supergravity; including the case of 4d stationary, axisymmetric gravity which we focus on in this paper. 
In other concurrent work on 6dCS theory \cite{Bittleston:2020hfv}, it was shown that 4dCS theory arises as a reduction of 6dCS theory, just as 2dIFTs arise as a reduction of the 4d sdYM equations. This established a commutative diagram relating Chern-Simons theories to integrable models. 
\begin{equation*}
\begin{tikzpicture}
\node at (-2,1) {\textbf{6dCS}};
\node at (-2,-1) {\textbf{4dCS}};
\node at (2,1) {\textbf{4dIFT}};
\node at (2,-1) {\textbf{2dIFT}};
\draw[->, very thick] (-1,1) -- (1,1);
\draw[->, very thick] (-1,-1) -- (1,-1);
\draw[->, very thick, decorate, decoration={snake, segment length=12.5pt, amplitude=2pt}] (-2,0.75) -- (-2,-0.75);
\draw[->, very thick, decorate, decoration={snake, segment length=12.5pt, amplitude=2pt}] (2,0.75) -- (2,-0.75);
\end{tikzpicture}
\end{equation*}
Localisation is represented in the diagram by straight arrows and dimensional reduction by squiggly arrows. 
In particular, since the two papers \cite{Bittleston:2020hfv} and \cite{Penna:2020uky} first appeared within days of one another, the 4dCS theory related to stationary, axisymmetric gravity was not explored in \cite{Penna:2020uky}. The present work seeks to fill this gap in the literature.

As mentioned earlier, our reduction of 6dCS theory results in a 4dCS theory with some novel features. These features of the 4dCS theory may be directly matched to novel features of the associated reduction. 
Firstly, we implement a discrete reduction which acts simultaneously on spacetime and on the gauge group. At the level of the 2d integrable model, this reduces the target space from the Lie group $G$ to the symmetric-space $G / G_{0}$ where $G_{0}$ is the fixed subgroup of a $\bZ_{2}$-automorphism. At the level of 4dCS theory, this produces the branch cut defect described above. To facilitate comparison with the literature, we isolate this ingredient in appendix \ref{app:symm_PCM} where we show how to recover the usual spacetime-independent branch cut defects in 4dCS theory from 6dCS theory.

Secondly, our reduction features a rotational vector field. 
Unlike the translational reductions considered in \cite{Bittleston:2020hfv}, it is known that rotational vectors have a non-trivial lift to twistor space (see e.g.\ \cite{Mason:1996, Penna:2020uky}), mixing the spectral parameter $\bCP^{1}$ with the spacetime $\bR^{4}$. 
When applied to 6dCS theory, this forces us to define a new invariant spectral parameter which necessarily depends on the spacetime coordinates. Following this through to 4dCS theory, we find the spacetime-dependent 1-form $\omega$ discussed above. At the level of the 2dIFT, rather than landing on the standard symmetric-space $\sigma$-model, the action comes with a spacetime-dependent coupling parameter which breaks some of the spacetime symmetries. Remarkably, this model is nonetheless integrable (see \cite{Hoare:2020fye} for recent work on integrability-preserving spacetime-dependent couplings in more general $\sigma$-models).

\subsection{Summary of contents}

Let us give a short summary of the contents of this paper. In section \ref{sec:background}, we review perhaps the most famous example of integrability in gravity: stationary, axisymmetric vacuum solutions in GR. Both the Lax formalism and reduction to a 2d $\sigma$-model are presented. The remainder of the paper will be devoted to establishing this 2dIFT in the commutative diagram of models shown above.  
Section \ref{sec:4dCS} is intended to be self-contained, and describes the relationship between 4dCS theory and integrability in gravity. We show in detail how the sigma-model for axisymmetric gravity is recovered from 4dCS theory via localisation of the action to 2d defects. The reader primarily interested in this key result could focus on this section. 
Section \ref{sec:4dWZW} reviews how such an integrable 2d $\sigma$-model can alternatively be derived as a reduction of the 4d Wess-Zumino-Witten (WZW) model. For our purposes, the 4dWZW model serves as a stepping stone to 6dCS theory. In fact, our 4dCS model was constructed by taking the reduction vectors for the right hand side of the diagram and lifting them to twistor space, so that we could apply the equivalent reduction to 6dCS theory. This lift is explained in section \ref{sec:twistorreduction}. We also provide a pedagogical introduction to 6dCS theory in the appendices. 
Finally, in section \ref{sec:6dCS_to_4dCS}, we present the reduction from 6dCS to 4dCS represented on the left hand side of the diagram. We conclude with an outlook on future work made possible by this new approach to integrability in gravity.

\section{Background on integrability in gravity}
\label{sec:background}

The sector of 4d vacuum General Relativity (GR) we are interested in consists of solutions with two commuting Killing vectors. These metrics can be written in the form
\begin{equation}\label{4d_metric}
    \dr s_4^2=e^{2\nu}(\dr \rho^2 +\dr z^2)+ \rho \, G_{mn}\dr x^m \dr x^n, \qquad m,n \in \{3,4\}
\end{equation}
where the function $\nu$ and the matrix $G$ depend only on the Weyl canonical coordinates $(\rho,z)$, and
\begin{equation}
    \text{det} \, G = \epsilon, \qquad \epsilon =\pm 1 \ .
\end{equation}
For the sake of concreteness, let us consider \textit{stationary} and \textit{axisymmetric} spacetimes ($\epsilon = -1$), for which we identify the pair of commuting killing vectors with $\partial_t$ and a vector for the azimuthal symmetry, $\partial_\phi$. This includes the Schwarzschild and Kerr black holes as well as the Kerr-NUT solution. 
It is a particularly well-studied sector of the theory, thanks in part to the solution generating techniques developed by Belinsky and Zakharov in the 1970s \cite{Belinsky:1971nt,Belinsky:1979mh}. Based on inverse scattering methods, their technique allows for an infinite number of solutions to be constructed from a given `seed' solution to the Einstein equations. 

The reason these solution generating techniques are possible is that Einstein's equations are integrable when specialised to stationary and axisymmetric spacetimes. Substituting the metric ansatz \eqref{4d_metric} into Einstein's equations, one finds that they decompose into a set of equations for $G$ and another for $\nu$ (given $G$). Defining a pair of $2 \times 2$ matrices $U_\rho$, $U_z$ by
\begin{equation}
     U_\rho \equiv \rho \, \partial_\rho G ~ G^{-1} \ , 
     \qquad U_z \equiv \rho \, \partial_z G ~ G^{-1} \ ,
\end{equation}
the vacuum Einstein equations can be written
\begin{align}\label{EE}
    \partial_\rho U_\rho +\partial_z U_z &=0 \ , 
    \\
    \partial_\rho \nu = \frac{1}{8\rho} \text{tr}(U_\rho^2-U_z^2)-\frac{1}{2\rho} \ , & \quad \partial_z \nu=\frac{1}{4\rho} \text{tr}(U_\rho \, U_z) \ . \label{EE_nu}
\end{align}
These equations define a completely integrable system --- a rare thing, especially in theories of gravity. The algebraic structure behind this integrability was first explored by Geroch, Breitenlohner, Maison, and others \cite{Geroch:1970nt, Geroch:1972yt, Breitenlohner:1986um}.

The integrability of this system is exhibited by the existence of a flat connection, called the Lax connection. The Lax for stationary and axisymmetric GR can be succinctly written in complex coordinates $\xi=z+i \rho$, $\Bar{\xi}=z-i \rho$ as the one-form 
\begin{equation}\label{eq:GR_lax}
    \mathcal{L}=\frac{- \pd_{\xi} G G^{-1}}{1 - \rmi Z}  \dr \xi+\frac{- \pd_{\bar{\xi}} G G^{-1}}{1 + \rmi Z} \dr \Bar{\xi}, \qquad Z =\frac{2 \rmi}{\xi-\bar{\xi}}\left(\frac{\xi+\bar{\xi}}{2}-W \pm \sqrt{(W-\xi)(W-\bar{\xi})}\right)
\end{equation}
where $Z$ and $W$ are complex parameters known in the literature as the variable and constant spectral parameters, respectively. Readers familiar with the Principal Chiral Model (PCM) will note that its Lax connection would be identical were we to neglect the spacetime dependence of the variable spectral parameter $Z$. The flatness of this connection for any value of $W$ is equivalent to the Einstein equations for $G$,
\begin{equation}
    \partial_\xi \mathcal{L}_{\Bar{\xi}}-\partial_{\Bar{\xi}} \mathcal{L}_\xi +[ \mathcal{L}_\xi,\mathcal{L}_{\Bar{\xi}}]=0 \quad \forall \, W \quad \Longleftrightarrow \quad \text{Eq. \eqref{EE}}.
\end{equation}
Since the flatness of a connection implies the existence of solutions $\Psi(\rho,z,W)$ to the equations
\begin{equation}
    \nabla_\xi \Psi=0, \qquad \nabla_{\bar{\xi}} \Psi=0 \ , \qquad \nabla 
    \equiv \dr + \mathcal{L} \ ,
\end{equation}
the Einstein equations for $G$ are sometimes presented as the compatibility conditions for this pair of linear `Lax equations'. Some authors work in a basis where $Z$ (rather than $W$) is held fixed, in which case the differential operators appearing above are mapped to 
\begin{align}
    \partial_\xi \hspace{0.5em} \mapsto \hspace{0.5em} \partial_\xi +(\xi-\bar{\xi})^{-1} \frac{1+\rmi Z}{1-\rmi Z} ~ Z \partial_Z \ , \qquad 
    \partial_{\bar{\xi}} \hspace{0.5em} \mapsto \hspace{0.5em} \partial_{\bar{\xi}} -(\xi-\bar{\xi})^{-1} \frac{1-\rmi Z}{1+\rmi Z} ~ Z \partial_Z \ . 
\end{align}
For example, in the original reference \cite{Belinsky:1979mh} the spectral parameter $\lambda$ held fixed is related those appearing here as $\lambda= - \rho Z$, in terms of which the Lax equations are
\begin{align}
    \left(\partial_z-\frac{2\lambda^2}{\lambda^2+\rho^2}\partial_\lambda\right) \Psi=\frac{\rho U_z-\lambda U_\rho}{\lambda^2+\rho^2} \,  \Psi, \qquad \left(\partial_\rho+\frac{2\lambda \rho}{\lambda^2+\rho^2}\partial_\lambda\right) \Psi=\frac{\rho U_\rho +\lambda U_z}{\lambda^2+\rho^2} \, \Psi \ . 
\end{align}

The effectively two-dimensional character of stationary, axisymmetric GR can be made explicit at the level of the action. 
We can see this schematically by evaluating the four-dimensional Einstein-Hilbert term for the metric \eqref{4d_metric},
\begin{equation}\label{schematic_sigma_model}
    \sqrt{\text{det} g^{(4)}} ~ R^{(4)} 
    = - ~ \frac{ \rho}{4} ~  \text{tr} [(G^{-1} \partial_\rho G)^2+(G^{-1} \partial_z G)^2]+\mathcal{B} \ , 
\end{equation}
where $\mathcal{B}$ collects various total derivatives. 
Given our determinant constraint on $G$, and the fact that it should be a symmetric matrix, the target space for 
the 2d model should be SL$(2, \mathbb{R})/$SO(2). In the next subsection, we carefully go through the reduction procedure of 4d GR by a pair of Killing vectors to come to the same basic conclusion. The reader uninterested in these details can skip ahead to section \ref{sec:4dCS}, where we present a new framework for understanding this 2d model for stationary and axisymmetric GR.

\subsection{Reduction of 4dGR to 2dIFT}

In this subsection, we consider the dimensional reduction of pure Einstein gravity to 2d by a pair of commuting Killing vectors, $\partial_\phi$ and $\partial_t$. For convenience, we work with Euclidean metric signature ($\epsilon=+1$) although results can easily be extended to Lorentzian signature.
The resulting nonlinear, coset space $\sigma$-model is sometimes called the BM model in the literature, after \cite{Breitenlohner:1986um}. We roughly follow the presentation of the reduction in that early work. 

Let us distinguish indices on the two isometry directions with lowercase Latin letters, and indices on the pair of associated orbit space directions with lowercase Greek letters,
\begin{align}
x^M=(x^\mu, x^{ m }), \qquad \mu=1,2, \qquad  m =3,4,
\end{align}
having identified $\partial_\phi$, $\partial_t$ with $\partial_3$, $\partial_4$. We will use the monikers `external' for $x^1$, $x^2$ and `internal' for $x^3=\phi$, $x^4=t$.
The first step is to use local Lorentz transformations to select a lower-triangular 4d vielbein $E_M^A$, 
\begin{align}\label{4d_vielbein}
 E_M^A = \begin{pmatrix}
\sqrt{\lambda} ~ e^{\alpha}_\mu & 0  \\
\sqrt{\rho} ~ \hat{e}^{ a }_{ n } ~ B^{n}_\mu  & \sqrt{\rho} ~ \hat{e}^{ a }_{ m }
\end{pmatrix}, \qquad g_{MN}^{(4)}=\eta_{AB} E^A_M E^B_N   
\end{align}
The explicit conformal factors $\lambda(x^1,x^2)$, $\rho(x^1,x^2)$ have been included so that we can normalize the determinants of the `purely external' vielbein $e$ and `purely internal' vielbein $\hat{e}$ as desired. Let us select 
\begin{equation}
\rho = \text{det } E^{ a }_{ m } \quad \Longleftrightarrow \quad \text{det } \hat{e}^{ a }_{ m } =1.
\end{equation}
Given some basis of coordinate one-forms, we may want to contract \eqref{4d_vielbein} by $\dr x^M$ and write
\begin{align}
E^{\alpha}= \sqrt{\lambda} ~ e^{\alpha}, \quad E^{ a }= \sqrt{\rho} ~ (\hat{e}^{ a }+ \hat{e}^{ a }_{ n } ~ B^{n} ), \quad \dr s_4^2=\eta_{\alpha \beta} E^{\alpha} E^{\beta} +\eta_{ a  b  } E^{ a } E^{ b  } \ , 
\end{align}
where the one-forms $e^{\alpha}$, $B^{n}$ have legs only on $\dr x^1$, $\dr x^2$ while $\hat{e}^{ a }$ has legs only on $\dr \phi$, $\dr t$. 
This lower-triangular ansatz for the vielbein is preserved under 2d Lorentz transformations acting on only $e^{\alpha}_\mu$ or only $\hat{e}^{ a }_{ m } $, as well as any 2d diffeomorphisms on $\hat{e}^{ a }_{ m } $. The only other diffeomorphisms which preserve the lower-triangular form are GL$(2,\mathbb{R})$ transformations on the internal coordinates $x^{ m }=(\phi,t)$ and those of the form
\begin{equation}
x^{ m } \mapsto x^{ m }+\Gamma^{ m }, \qquad B^{n} \mapsto B^{n} + \dr \Gamma^{ n }.
\end{equation}
From the Kaluza-Klein perspective $B^{n}$ is a pair of gauge fields in 2d, with field strengths
\begin{equation}
    F_{\mu \nu}^{n}=\partial_\mu B_\nu^{n}-\partial_\nu B_\mu^{n}.
\end{equation}
However, they are non-dynamical in 2d. To see this, define the 2d metric $g$ and associated Ricci scalar with respect to $e^\alpha_\mu$, and construct an analogous matrix $G$ from $\hat{e}^a_m$, 
\begin{align} \label{eq:MetricMatrices}
   g^{(2)}_{\mu \nu}= \eta_{\alpha \beta} e^{\alpha}_\mu e^{\beta}_\nu, \qquad G_{ m  n }  = \eta_{ a  b  } \hat{e}^{ a }_{ m } \hat{e}^{ b  }_{ n }.
\end{align}
Suppressing internal indices $ m , n $,\footnote{So e.g. $G F_{\mu \nu}=G_{mn} F_{\mu \nu}^{n}$ } the 4d Einstein-Hilbert term can then be written
\begin{equation} \label{EH_to_2d}
\begin{aligned}
    \sqrt{ \text{det } g^{(4)}} ~ R^{(4)} =  \rho \sqrt{ \text{det } g^{(2)}} & ~ \bigg[ R^{(2)}
    - \frac{1}{4} \text{tr} (G^{-1} \partial_\mu G ~G^{-1} \partial^\mu G)
    \\
    & \qquad + \frac{1}{4\lambda} \rho ~ F_{\mu \nu}^T ~ G ~ F^{\mu \nu}
    + \lambda^{-1} \partial_\mu \lambda ~ \rho^{-1} \partial^\mu \rho
    \bigg] \ , 
\end{aligned}
\end{equation}
with all $\mu,\nu$ indices contracted using the 2d metric $g^{(2)}_{\mu \nu}$.
The corresponding field equations for the $B_\mu^{n}$ and its scalar dual $F_0$ read
\begin{equation}
    \nabla_\mu (\rho^2 \lambda^{-1} G ~ F^{\mu \nu})=0, \qquad 
    \partial_\mu F_0=0.
\end{equation}
We see that the equations of motion fix $F_0=$ const.\ which means $F_0$ and $F_{\mu \nu}$ must both be set to zero for asymptotically flat solutions. As we are primarily interested in this class of solutions, the $F_{\mu \nu}^T ~ G ~ F^{\mu \nu}$ term will be dropped from the Lagrangian moving forward. 

The remaining equations of motion are
\begin{align}
\begin{split}
    R_{\mu \nu}^{(2)}-\frac{1}{2}g_{\mu \nu}^{(2)} R^{(2)} =  ~ & \frac{1}{4} \text{tr} (G^{-1} \partial_\mu G ~G^{-1}  \partial_\nu G) -\lambda^{-1} \partial_{(\mu} \lambda ~ \rho^{-1} \partial_{\nu)}\rho
    \label{2d_tensor_eq}\\
    & -\frac{1}{2} g_{\mu \nu}^{(2)} \left[\frac{1}{4} \text{tr} (G^{-1}  \partial_\sigma G ~G^{-1} \partial^\sigma G) -\lambda^{-1} \partial_{\sigma} \lambda ~ \rho^{-1} \partial^{\sigma}\rho\right] \ , 
\end{split} \\
    \nabla_\mu (\rho ~ G^{-1} \partial^\mu G) = ~ & 0 \ , 
    \label{2d_vector_eq}\\
    \nabla_\mu \partial^\mu \rho = ~ & 0 \ . 
    \label{2d_scalar_eq}
\end{align}
With a suitable choice of coordinates and redefinition of $\lambda$, we can bring $g_{\mu \nu}^{(2)}$ to $\delta_{\mu \nu}$, setting the Ricci tensor and Ricci scalar to zero and replacing all covariant derivatives with partials in the equations above. The scalar equation $\pd_{\mu} \partial^{\mu} \rho = 0$ says that $\rho$ is a harmonic function on $\mathbb{R}^2$. Together with it's conjugate harmonic function $z$, defined by $\dr z= -\star_2 \dr \rho$, it supplies us with canonical coordinates $(\rho,z)$ in 2d. 
The flat-space equation for $G$ then reads 
\begin{equation}\label{G_eq}
     \partial_\mu (\rho ~ G^{-1} \partial^\mu G)=0 \ ,
\end{equation}
while the field equations \eqref{2d_tensor_eq} become
\begin{align}\label{lambda_eqs}
    \partial_z \log \lambda= \frac{\rho}{2} \text{tr} (G^{-1} \partial_\rho G ~G^{-1}  \partial_z G), \qquad \partial_\rho \log \lambda = \frac{\rho}{4} \tr [(G^{-1} \partial_\rho G)^2-(G^{-1} \partial_z G)^2] \ .
\end{align}
These can be interpreted as fixing $\lambda$ up to a single integration constant. The integrability conditions for these equations are ensured by \eqref{G_eq}. 

In summary, the field content of the 2d theory is a $2 \times 2$ matrix $G$, two abelian gauge fields $B^{n}$, a scalar field $\lambda$, and 2d gravity. When it comes to the local dynamics of this theory, the gauge fields must be trivial and the 2d metric is flat up to diffeomorphism. Furthermore, the scalar field $\lambda$ is determined by \eqref{lambda_eqs} given a solution $G$ to \eqref{G_eq}. 
Therefore, finding a local solution to 4d GR amounts to finding a solution to the flat-space equation for $G$. Notably, this is only a local statement at the level of equations of motion. A more thorough treatment may involve topological contributions from both the gauge theory and gravitational sectors. 

If we are interested in finding classical solutions to 4d gravity with these assumptions, then it is sufficient to solve the flat-space equation \eqref{G_eq} for $G$. This is none other than the equation of motion for the nonlinear $\sigma$-model with action 
\begin{equation}\label{2d_sigma}
    S=  -\frac{1}{4} \int \dr \rho ~ \dr z \, 
   \rho \, \text{tr} [(G^{-1} \partial_\rho G)^2+(G^{-1} \partial_z G)^2].
\end{equation}
Indeed, bringing the 2d metric to $\delta_{\mu \nu}$, this is the second term in the reduced GR action \eqref{EH_to_2d} (the first and third term having been turned off). 
Crucially, the $\sigma$-model field $G$ is built from the vielbein on the isometry directions. We can use this to determine the degrees of freedom contained in $G$. The vielbein is a $2 \times 2$ matrix, and we have a chosen a normalisation such that it has unit determinant. This means that $\hat{e}$ lives in SL$(2,\bR)$, but the metric $G$ only depends on $\hat{e}$ through the combination \eqref{eq:MetricMatrices}. In particular, the metric is invariant under SO$(2)$ transformations of the vielbein, meaning that the action \eqref{2d_sigma} describes\footnote{For Lorentzian signature, the metric is invariant under SO$(1,1)$ transformations of the vielbein so the coset $\sigma$-model is SL$(2,\bR)/$SO$(1,1)$.} an SL$(2,\bR)/$SO$(2)$ coset $\sigma$-model. We would like to highlight that this is not the usual SL$(2,\bR)/$SO$(2)$ coset $\sigma$-model due to the spacetime-dependent coupling parameter appearing in the action.

Looking back at the 4d metric ansatz in \eqref{4d_metric}, we see how it is adapted to this integrable structure. Identifying $\lambda = \rho e^{2\nu}$ completes the match between the equations of motion in \eqref{lambda_eqs} and \eqref{G_eq}, on the one hand, and \eqref{EE} and \eqref{EE_nu} on the other. The key takeaway from this section for the remainder of this work is that the 2d $\sigma$-model given in \eqref{2d_sigma} emerges from the reduction of pure 4d GR on a pair of isometry directions, such that solutions to the full 4d Einstein equations can be constructed given a solution for the $\sigma$-model field $G$. Integrating \eqref{lambda_eqs} to obtain $\lambda$ is obviously an important final step in these solution generating techniques. But for our present purposes, the $\sigma$-model term \eqref{2d_sigma} alone will serve as the bridge between the actions for 4d GR and the Chern-Simons models we are interested in.

\section{4dCS to 2dIFT} \label{sec:4dCS}

\begin{equation*}
\begin{tikzpicture}
\node at (-2,1) {\textcolor{tab10lightgray}{\textbf{6dCS}}};
\node at (-2,-1) {\textbf{4dCS}};
\node at (2,1) {\textcolor{tab10lightgray}{\textbf{4dIFT}}};
\node at (2,-1) {\textbf{2dIFT}};
\draw[->, draw=tab10lightgray, very thick] (-1,1) -- (1,1);
\draw[->, draw=tab10red, very thick] (-1,-1) -- (1,-1);
\draw[->, draw=tab10lightgray, very thick, decorate, decoration={snake, segment length=12.5pt, amplitude=2pt}] (-2,0.75) -- (-2,-0.75);
\draw[->, draw=tab10lightgray, very thick, decorate, decoration={snake, segment length=12.5pt, amplitude=2pt}] (2,0.75) -- (2,-0.75);
\end{tikzpicture}
\end{equation*}

Recent developments in the field of two-dimensional integrable models have brought to light the existence of four-dimensional Chern-Simons (4dCS) theory \cite{Costello:2019tri}. 
This higher dimensional gauge theory provides a geometric origin for the spectral parameter appearing in the Lax formalism, making manifest the integrable structure of the lower dimensional model. 
The 4-manifold $M_{4}$ over which it is defined is the product of the spectral plane $\bCP^{1}$ and the 2d spacetime of the integrable model $\bR^{2}$. 
The action is built from the usual Chern-Simons 3-form for an algebra-valued gauge field $A$, and a meromorphic 1-form $\omega$ with poles at certain values of $Z \in \bCP^{1}$. 
This action is given by 
\begin{equation}\label{4dCS_action}
    S_{\text{4dCS}} = \frac{1}{2 \pi \rmi} \int_{M_{4}} \omega \wedge \tr \bigg( A \wedge \dr A + \frac{2}{3} A \wedge A \wedge A \bigg) \ . 
\end{equation}
The poles as well as the zeros of $\omega$ constitute the essential data of the theory.
Before introducing the specific $\omega$ relevant for stationary and axisymmetric gravity, let us take a moment to outline how 4dCS theories localize to 2d field theories in general. We refer the reader to \cite{Lacroix:2021iit} for a pedagogical introduction to the topic.

In three dimensions, Chern-Simons theory is a topological field theory. This statement may be understood by considering the local degrees of freedom in the fundamental gauge field $A$. 
The equation of motion $F = 0$ implies that there exists a local solution $g$ to the equation $A = g^{-1} \dr g$. 
Any gauge field of this form may be fixed to $A = 0$ by a gauge transformation, so all solutions to the equations of motion are locally gauge trivial. 
The fact that 3dCS theory has no local degrees of freedom is a manifestation of its topological nature. 

However, let us consider 3dCS theory on a manifold with boundary. 
We must impose boundary conditions on both the fundamental field $A$ and on the gauge transformations. 
In effect, the presence of a boundary has broken some of the gauge symmetry of the theory. 
It may no longer be possible to transform a gauge field $A = g^{-1} \dr g$ into $A = 0$ since the required gauge transformation may not satisfy the boundary conditions. 
In the bulk of the manifold (away from the boundary, that is) this has no effect and there are still no local dynamics. 
On the other hand, the broken gauge symmetry gives rise to local degrees of freedom which live on the boundary.

The same argument applies to 4dCS theory with the role of the boundary played by the poles in $\omega$. 
It is necessary to impose boundary conditions on the gauge transformations at these points, and this leads to the emergence of physical degrees of freedom living on $\bR^{2}$ alone. 
We will refer to these degrees of freedom as \textit{edge modes}, and they will become the fundamental field $G$ of the 2d theory -- an SL$(2, \mathbb{R})/$SO(2) $\sigma$-model, for our present purposes.

\subsection{Constructing our 4dCS theory}

We will continue to work with coordinates $\{ \rho , z \}$ on spacetime and consider the 4dCS theory \eqref{4dCS_action} with meromorphic 1-form
\begin{equation}\label{eq:our_omega}
    \omega = -\frac{\rho}{2} \bigg( \frac{Z^{2} + 1}{Z^{2}} \, \dr Z \bigg) + \frac{Z^{-1} - Z}{2} \, \dr \rho + \dr z \ . 
\end{equation}
This may seem like an ad hoc choice. For the time being, we will adopt the philosophy that ``the proof is in the pudding'', and justify this by showing that it leads to the 2d integrable model known to describe stationary axisymmetric GR. 
However, in section \ref{sec:6dCS_to_4dCS} we will show that \eqref{eq:our_omega} is a consequence of the $\partial_\phi$, $\partial_\tau$ reduction isometries, emerging from a corresponding reduction of 6dCS theory to our 4dCS model.

To the best of our knowledge, all previous 4dCS constructions have considered 1-forms $\omega$ which may be written in terms of a twist function $\varphi (Z)$ as $\omega = \varphi(Z) \, \dr Z$. In particular, they only have legs on $\bCP^{1}$ and only depend on the $\bCP^{1}$ directions. By comparison, our meromorphic 1-form $\omega$ mixes the spacetime and $\bCP^{1}$ directions. We will now describe the construction of this 1-form by starting with a more conventional 1-form and introducing a spacetime-dependent branch cut. 
Consider another spectral plane parameterised by $W \in \bCP^{1}$ and equipped with the meromorphic 1-form $\dr W$. 
This 1-form has a second order pole at $W = \infty$, which can be seen by moving to the other patch covering $\bCP^{1}$. 
Now, let us insert a branch cut in this spectral plane between the points $W = z + \rmi \rho$ and $W = z - \rmi \rho$. 
This introduces a two-sheeted covering of the $W$-plane which we can parameterise by the coordinate 
\begin{equation}
    Z = \frac{1}{\rho} \Big( z - W \pm \sqrt{(W - z)^{2} + \rho^{2}} \Big) \ . 
\end{equation}
Since this is a double covering, there are two values of $Z$ for each value of $W$ except for the points at the end of the branch cut where the radicand vanishes. 
We can invert this relation to find the two-to-one map from $Z$ to $W$, 
\begin{equation}
    W = z + \frac{\rho}{2} \big( Z^{-1} - Z \big) \ . 
\end{equation}
We can move between the two sheets of the double covering with the map $Z \mapsto -Z^{-1}$, which preserves $W$. 
The meromorphic 1-form $\omega$ given in \eqref{eq:our_omega} is simply $\dr W$ after moving to the two-sheeted cover parameterised by $Z$. 
This $\omega$ has two second order poles, at $Z = 0$ and $Z = \infty$, which are the preimages of $W = \infty$. 

Despite our unconventional choice of 1-form, the essential properties of the 4dCS theory are retained. Since this is a surprising result, we will take some time to demonstrate this statement and review the fundamentals of 4dCS theory. It is convenient to work in complex coordinates on spacetime given by $\xi = z + \rmi \rho$ and $\bar{\xi} = z - \rmi \rho$. 
In these coordinates, the meromorphic 1-form is given by 
\begin{equation}\label{eq:omega_of_Z}
    \omega = \frac{\rmi (\xi - \bar{\xi})}{4} \bigg( \frac{Z^{2} + 1}{Z^{2}} \, \dr Z \bigg) + \frac{\rmi (Z - \rmi)^{2}}{4 Z} \, \dr \xi - \frac{\rmi (Z + \rmi)^{2}}{4 Z} \, \dr \bar{\xi} \ . 
\end{equation}
The singularities in $\omega$ play the role of boundaries in our theory and we must impose boundary conditions on the gauge field $A$ at these points. To see this, consider the variation of the action, 
\begin{equation}
    \delta S_{\text{4dCS}} = \frac{2}{2 \pi \rmi} \int_{M_{4}} \omega \wedge \tr \big( \delta A \wedge F \big) + \frac{1}{2 \pi \rmi} \int_{M_{4}} \dr \omega \wedge \tr \big( \delta A \wedge A \big) \ . 
\end{equation}
The first term gives the bulk equations of motion $\omega \wedge F = 0$, while the second term is a boundary term which must be killed by imposing constraints on the gauge field. 
One might expect that $\dr \omega$ is identically zero since the 1-form may be expressed as $\omega = \dr W$. 
This is almost correct, but the argument fails where $\omega$ is singular. 
At these points, we must make use of the identities from complex analysis 
\begin{equation}
    \pd_{\bar{Z}} \bigg( \frac{1}{Z} \bigg) = - 2 \pi \rmi \, \delta (Z) \ , \qquad 
    \int_{\bCP^{1}} \dr Z \wedge \dr \bar{Z} \, \delta(Z) \, f(Z) = f(0) \ . 
\end{equation}
This means that $\dr \omega$ is a distribution with support on $\bCP^{1}$ at the poles of $\omega$. 
It is for this reason that we refer to the second term in the variation as a boundary term. More explicitly, we can write the contribution from $Z=0$ as 
\begin{align}\label{eq:domega}
    \frac{\dr \omega}{2\pi \rmi} = 
    \partial_{Z} \delta(Z) ~  \frac{\rmi (\xi - \bar{\xi})}{4} \dr \bar{Z} \wedge \dr Z + \delta (Z) \frac{\rmi}{4 } \, \dr \bar{Z} \wedge \dr \xi -\delta (Z) \frac{\rmi}{4 } \, \dr \bar{Z} \wedge \dr \bar{\xi}  .
\end{align}
Sufficient boundary conditions on the gauge field which cause the boundary term in $\delta S_{\text{4dCS}}$ to vanish are given by\footnote{When we write the boundary conditions on the gauge field, we would like to highlight that the $\dr Z$ legs do not contribute due to the restriction map, that is $\dr Z \vert_{Z = 0} = 0$. This is in contrast to the boundary variation where the $Z$-component of $A$ will appear explicitly. Despite this observation, it is sufficient to constrain only the $\xi$ and $\bar{\xi}$-components in this basis.} 
\begin{equation} \label{eq:fieldBC}
    A \big\vert_{Z = 0} = 0 \ , \qquad
    A \big\vert_{Z = \infty} = 0 \ . 
\end{equation}
By a similar argument, the gauge transformations $\delta_\epsilon A = \dr \epsilon + [A , \epsilon]$ must obey boundary conditions given by 
\begin{equation} \label{eq:4dgaugeBC}
    \dr \epsilon \big\vert_{Z = 0} = 0 \ , \qquad
    \dr \epsilon \big\vert_{Z = \infty} = 0 \ . 
\end{equation}

The zeroes of $\omega$ also play an important role in 4dCS theory. Due to the zeros at $Z = \pm \rmi$, we should allow the gauge field $A$ to have simple poles at these points. If $A_\xi$ has a simple pole at $Z = \rmi$ and $A_{\bar{\xi}}$ has a simple pole at $Z = -\rmi$, the action $S_\text{4dCS}$ remains finite. The presence of singularities in the gauge field is far from a problem in 4dCS theory -- it is a crucial feature to capture the usual meromorphic dependence of a Lax connection. We will therefore allow these singularities in our field configurations. 

Turning to the symmetries of the action, notice that we have a trivial shift symmetry acting as 
\begin{equation}
    A \mapsto A + C_{\omega} \, \omega \ . 
\end{equation}
In the 1-form basis  $\{ \dr W , \dr \bar{W} , \dr \xi , \dr \bar{\xi} \}$ adapted to this symmetry, $\omega=\dr W$ and thus the component $A_W$ does not contribute to the action -- it decouples. For this reason we will often neglect to mention $A_W$ as it can always be fixed to zero. Transforming to the basis $\{ \dr Z , \dr \bar{Z} , \dr \xi , \dr \bar{\xi} \}$, this choice also sets $A_{Z} = 0$.

This almost completes our definition of the theory, but there is one final ingredient which we would like to introduce. 
The interpretation of the $Z$-plane as a double covering of the $W$-plane calls for an additional restriction on the gauge field $A$. 
Rather than having generic dependence on $Z$, we would like to think of the gauge field $A$ as living on the spectral plane parameterised by $W$. 
One might think to impose the constraint $A(Z) = A(-1/Z)$ so that $A$ is single-valued on the $W$-plane. We instead allow a non-trivial transformation on the Lie algebra indices of $A$, introducing the $\bZ_{2}$-automorphism of the algebra $\fg$. 
In the context of stationary axisymmetric gravity, the appropriate Lie algebra is $\fg = \mathfrak{sl}(2,\bR)$ and the automorphism $\eta : \fg \to \fg$ is given by $x \mapsto - x^{T}$. 
We impose the equivariance condition 
\begin{equation} \label{eq:equivariance}
    A(Z) = \eta \big( A(-1/Z) \big) \ . 
\end{equation}
This means that the values of the gauge field on each sheet of the covering are not independent, and knowing one is sufficient to know them both. 
In the resulting 2dIFT, this restricts the target space from SL$(2,\bR)$ to the coset $\text{SL}(2,\bR)/\text{SO}(2)$. 
It is also important that this is compatible with the boundary conditions we have imposed on the gauge field. 
In the literature on 4dCS theory, this equivariance condition is also known as a \textit{branch cut defect} \cite{Costello:2019tri}. 
This completes the definition of our 4dCS theory.

\subsection{Localisation to 2dIFT}

Next, we would like to show that 4dCS theory localises to an integrable field theory on spacetime. 
In particular, unlike a Kaluza-Klein reduction in which infinitely many modes are discarded from the theory, a finite number of fields on $\bR^{2}$ capture all of the physical degrees of freedom in 4dCS theory.

As a first step in this localisation, it is helpful to introduce a field redefinition which separates the bulk gauge field (i.e. away from $Z=0$ or $Z=\infty$) from the edge modes. 
We introduce two new fields $\cL$ and $\hat{g}$ defined by 
\begin{equation}\label{eq:field_redef}
    A = \cL^{\hat{g}}, \qquad  \cL^{\hat{g}} \equiv \hat{g}^{-1} \cL \hat{g} + \hat{g}^{-1} \dr \hat{g} \ . 
\end{equation}
This new parameterisation is partially redundant and introduces an \textit{internal} gauge symmetry which acts as 
\begin{equation}
    \cL \mapsto \cL^{\check{h}} \ , \qquad 
    \hat{g} \mapsto \check{h}^{-1} \hat{g} \ . 
\end{equation}
Notice that the original field $A$ is invariant under this transformation, meaning that it preserves the action. We can leverage this internal symmetry to significantly simplify the equations of motion for the theory.
Denoting the contraction of a vector field $X$ with a differential form $A$ by $X \vee A$, we will impose the gauge fixing constraint 
\begin{equation}\label{eq:int_gauge}
    \pd_{\bar{Z}} \vee \cL = 0 \quad \Longleftrightarrow \quad 
    \cL_{\bar{Z}} = 0 \ . 
\end{equation}
There are some residual symmetries after this gauge fixing, including the portion of the original gauge symmetry satisfying the boundary conditions. 
In a moment, we will use the remaining gauge symmetries to impose constraints on the edge mode $\hat{g}$, but first let us return to the action. 
In the new variables, it is given by 
\begin{equation}
    S_{\text{4dCS}} = 
    \frac{1}{2 \pi \rmi} \int_{M_{4}} \omega \wedge \tr \big( \cL \wedge \dr \cL \big) 
    + \frac{1}{2 \pi \rmi} \int_{M_{4}} \dr \omega \wedge \tr \big( \cL \wedge \dr \hat{g} \hat{g}^{-1} \big) 
    - \frac{1}{6 \pi \rmi} \int_{M_{4}} \omega \wedge \tr \big( \hat{g}^{-1} \dr \hat{g} \big)^{3} \ . 
\end{equation}
In the second term, the edge mode $\hat{g}$ appears against the 2-form $\dr \omega$. 
While it might appear that this term depends on the value of $\hat{g}$ over all of $M_{4}$, the presence of the distribution $\dr \omega$ means that it only depends on the fields 
\begin{equation}\label{eq:edge_modes}
    \hat{g} \big\vert_{Z = 0} = g \ , \qquad 
    \hat{g}^{-1} \pd_{Z} \hat{g} \big\vert_{Z = 0} = \phi \ , \qquad 
    \hat{g} \big\vert_{Z = \infty} = \tilde{g} \ , \qquad 
    \hat{g}^{-1} \pd_{Z} \hat{g} \big\vert_{Z = \infty} = \tilde{\phi} \ . 
\end{equation}
The $\dr \omega$ term cares about both the edge mode and its $\bCP^{1}$-derivatives because $\omega$ contains second order poles. 
Higher order poles would lead to higher order derivatives contributing to the action. 
Readers familiar with the 2d Wess-Zumino-Witten (WZW) model might recognise the 3-form in the third term as a Wess-Zumino (WZ) term. 
In that context, despite being a 3-form integrated over a 3-manifold, the WZ term only produces 2d dynamics on the boundary. 
Based on the similarity, one might expect that this remains true in the present context, and this is indeed the case. 

First, consider an extension of $\hat{g}$ over the 5-manifold $M_{5} = M_{4} \times [0,1]$ which reproduces $\hat{g}$ on $M_{4} \times \{ 1 \}$ and is the trivial map on $M_{4} \times \{ 0 \}$. 
Denoting this extension by the same symbol $\hat{g}$ in an abuse of notation, we may write the WZ term as a surface integral over $M_{5}$, 
\begin{equation}
    \int_{M_{4}} \omega \wedge \tr \big( \hat{g}^{-1} \dr \hat{g} \wedge \hat{g}^{-1} \dr \hat{g} \wedge \hat{g}^{-1} \dr \hat{g} \big) = \int_{M_{5}} \dr \bigg[ \omega \wedge \tr \big( \hat{g}^{-1} \dr \hat{g} \wedge \hat{g}^{-1} \dr \hat{g} \wedge \hat{g}^{-1} \dr \hat{g} \big) \bigg] \ . 
\end{equation}
Since the $\hat{g}$-dependent 3-form in this expression is closed, the exterior derivative on the right hand side can only act on $\omega$, producing the desired distribution on $\bCP^{1}$. 
This allows us to write the action as 
\begin{equation}\label{eq:S_4dCS_prepped}
    S_{\text{4dCS}} = 
    \frac{1}{2 \pi \rmi} \int_{M_{4}} \omega \wedge \tr \big( \cL \wedge \dr \cL \big) 
    + \frac{1}{2 \pi \rmi} \int_{M_{4}} \dr \omega \wedge \bigg[ \tr \big( \cL \wedge \dr \hat{g} \hat{g}^{-1} \big) - \text{WZ} [\hat{g}] \bigg] \ , 
\end{equation}
where we define the WZ 2-form by 
\begin{equation}
    \text{WZ} [\hat{g}] = \frac{1}{3} \int_{[0,1]} \tr \big( \hat{g}^{-1} \dr \hat{g} \wedge \hat{g}^{-1} \dr \hat{g} \wedge \hat{g}^{-1} \dr \hat{g} \big) \ . 
\end{equation}
From this expression for the action, we see that the edge mode indeed appears in the action only through the `boundary' fields \eqref{eq:edge_modes}, justifying its name.

Let us consider the degrees of freedom in the edge mode more carefully. 
Earlier, we saw that the original gauge symmetries of 4dCS theory are constrained to obey the boundary conditions \eqref{eq:4dgaugeBC}. 
While these restrict the spacetime derivatives of allowed gauge transformations, their $\bCP^{1}$-derivatives are unconstrained, meaning that we can choose to fix $\phi = 0$ and $\tilde{\phi} = 0$. 
We should also consider the compatibility of our field configurations with the equivariance condition \eqref{eq:equivariance}. 
This condition exchanges the two poles of $\omega$, meaning that the values of the edge mode at these points must be related by $\tilde{g} = \eta (g)$. 
In summary, the physical degrees of freedom in the edge mode are captured by 
\begin{equation}\label{eq:ext_gauge}
    \hat{g} \big\vert_{Z = 0} = g \ , \qquad 
    \hat{g}^{-1} \pd_{Z} \hat{g} \big\vert_{Z = 0} = 0 \ , \qquad 
    \hat{g} \big\vert_{Z = \infty} = \eta(g) \ , \qquad 
    \hat{g}^{-1} \pd_{Z} \hat{g} \big\vert_{Z = \infty} = 0 \ . 
\end{equation}

Now that we have the degrees of freedom of our theory in hand, and have applied some helpful gauge fixing conditions, we will continue with the localisation to spacetime. 
The next step is to solve a subset of the equations of motion to explicitly fix the $\bCP^{1}$-dependence of $\cL$. Since the equations of motion read
\begin{equation}
    \omega \wedge F = 0 \ , \qquad 
    F \equiv \dr \cL + \frac{1}{2} [\cL,\cL] \ ,
\end{equation}
it is convenient to work in the 1-form basis  $\{ \dr W , \dr \bar{W} , \dr \xi , \dr \bar{\xi} \}$ and its duals vector basis\footnote{We use $\mathring{}$ to distinguish vector and form components expressed in this basis from those in the usual $Z, \bar{Z},\xi, \bar{\xi}$ coordinate basis.} $\{ \mathring{\partial}_W, \mathring{\partial}_{\bar{W}}, \mathring{\partial}_\xi, \mathring{\partial}_{\bar{\xi}}\}$, so that we have
\begin{equation}
    \dr W \wedge F =0 \quad \Leftrightarrow \quad \mathring{F}_{\xi \bar{\xi}}=0 \ , \quad \mathring{F}_{\bar{W} \xi}=\mathring{F}_{\bar{W} \bar{\xi}}=0 \ .
\end{equation}
The $\mathring{F}_{\xi \bar{\xi}}$ equation will become the equation of motion for the 2dIFT. Using the (internal) gauge-fixing \eqref{eq:int_gauge}, the latter two equalities tell us
\begin{equation}
    \mathring{\pd}_{\bar{W}} \mathring{\cL}_{\xi} = 0 \ , \qquad \mathring{\pd}_{\bar{W}} \mathring{\cL}_{\bar{\xi}} = 0 \ ,
\end{equation}
meaning that these components of $\cL$ are holomorphic functions of $W$. The relationship between the $\partial_{a}$ and $\mathring{\partial}_{a}$ bases ensure $\mathring{\cL}_{\xi}$ and $\mathring{\cL}_{\bar{\xi}}$ are holomorphic functions of $Z$ as well. If they were bounded, this would imply they were constant by Liouville's theorem. However, recall that we have allowed these components of our gauge field to have singularities: $\cL_{\xi}$ is allowed a simple pole at $Z = - \rmi$ and $\cL_{\bar{\xi}}$ is allowed a simple pole at $Z = +\rmi$. Moving to the other basis, the same can be said about the components $\mathring{\cL}_{\xi}$ and $\mathring{\cL}_{\bar{\xi}}$. The general solution with these properties is given by 
\begin{equation} \label{eq:L_of_UV}
    \mathring{\cL}_{\xi} = \frac{1}{1 - \rmi Z} \, U_{\xi} - \frac{\rmi Z}{1 - \rmi Z} V_{\xi} \ , \qquad 
    \mathring{\cL}_{\bar{\xi}} = \frac{1}{1 + \rmi Z} \, U_{\bar{\xi}} + \frac{\rmi Z}{1 + \rmi Z} V_{\bar{\xi}} \ ,
\end{equation}
where the $U$s and $V$s are functions of the spacetime coordinates $\xi$, $\bar{\xi}$ alone. Note, we could have expressed each component of $\cL$ as the sum of a term which is constant on $\bCP^{1}$ and a term which is singular. 
Instead we have chosen to collect sums and differences of those two terms 
such that each term vanishes at either $Z = 0$ or $Z = \infty$. 
This parameterisation makes it particularly easy to solve the boundary conditions \eqref{eq:fieldBC} which, together with the field redefinition \eqref{eq:field_redef}, yields
\begin{equation}\label{eq:UV_sol}
    U_{\xi} = - \pd_{\xi} g g^{-1} \ , \qquad 
    U_{\bar{\xi}} = - \pd_{\bar{\xi}} g g^{-1} \ , \qquad
    V_{\xi} = - \pd_{\xi} \tilde{g} \tilde{g}^{-1} \ , \qquad 
    V_{\bar{\xi}} = - \pd_{\bar{\xi}} \tilde{g} \tilde{g}^{-1} \ .
\end{equation}
Substituting these solutions back into the expression for the Lax connection \eqref{eq:L_of_UV}, it does not immediately agree with the expected Lax given in \eqref{eq:GR_lax}. However, a gauge transformation\footnote{One might object that the gauge transformation by $\tilde{g}$ is not a symmetry of the theory. It would fix the value of the edge mode at infinity which would break the equivariance condition $\tilde{g} = \eta (g)$. One resolution is to understand the gauge equivalence of the Lax connections as a formal statement at the level of equations of motion: the flatness conditions are equivalent. Alternatively, one might weaken the equivariance condition, demanding that it is respected only up to a gauge transformation. Both approaches seem valid at this level.} by $\tilde{g}$ brings the Lax derived from 4dCS theory into the desired form,
\begin{equation}
    \mathring{\cL}_\xi =  \frac{- \pd_{\xi} G G^{-1}}{1 - \rmi Z} \ , \qquad 
    \mathring{\cL}_{\bar{\xi}} = \frac{- \pd_{\bar{\xi}} G G^{-1}}{1 + \rmi Z} \ , \qquad
    G \equiv  \tilde{g}^{-1} g \ .
\end{equation}
This derivation of the Lax connection for the 2d integrable model is a standard feature of 4dCS theories. 
In the application to stationary axisymmetric GR, this $G$ is precisely the SL$(2, \mathbb{R})/$SO(2) field of section \ref{sec:background}, encoding the metric components along the isometry directions. 
Indeed, the equivariance condition \eqref{eq:equivariance} tells us that $\tilde{g} = g^{T}$ implying $G = g^{T} g$ and we see that the edge mode $g$ is identified with the vielbein $\hat{e}$. 

Having fixed the $\bCP^1$-dependence of $\mathcal{L}$, we can return to the action \eqref{eq:S_4dCS_prepped} and localise it to two-dimensional spacetime. The first integrand is proportional to $\omega \wedge \tr(\cL \wedge \mathring{\pd}_{\bar{W}} \cL)$ and thus vanishes on-shell. This follows from $\omega=\dr W$ and the internal gauge fixing $\cL_{\bar{Z}}=0$ (which also sets $\mathring{\cL}_{\bar{W}}=0$), requiring $\dr \cL$ to saturate the $\dr \bar{W}$ leg. This leaves the boundary contributions, namely those terms proportional to $\dr \omega$.

For our choice of $\omega$ and the gauge fixing conditions \eqref{eq:ext_gauge} the boundary WZ-term vanishes. 
Since $\hat{g}^{-1} \pd_Z \hat{g}$ has been fixed to zero at $Z=0$ and $Z=\infty$, $\dr \omega$ must contribute a $\dr Z$ leg in this term. Performing an integration by parts, the contribution from $Z=0$ is
\begin{equation}
     \frac{1}{2\pi \rmi}  \int_{M_{4}} \dr \omega \wedge \text{WZ} [\hat{g}] 
     = - \frac{\rmi}{4} \int_{M_{4}}  \dr \bar{Z} \wedge \dr Z \wedge \partial_Z (\text{WZ} [\hat{g}]) ~ \delta(Z) ~  (\xi - \bar{\xi}) \ , 
\end{equation}
which vanishes, again since $\hat{g}^{-1} \pd_Z \hat{g}=0$ at the boundaries. The contribution from this term at $Z=\infty$ also vanishes. It remains to compute the second term in the action, whose contribution at $Z=0$ is given by
\begin{equation}
\begin{aligned}
    \frac{1}{2\pi \rmi} \int_{M_{4}} & \dr \omega \wedge  \tr \big( \cL \wedge \dr \hat{g} \hat{g}^{-1} \big)
    = \frac{\rmi}{4}  \int_{M_4} \dr Z \wedge \dr \bar{Z} \wedge \dr \xi \wedge \dr \bar{\xi} ~ \delta(Z) (\xi -\bar{\xi})
     \\
     & \qquad \times  \tr \left[ \left(\partial_Z \cL_\xi+(\xi-\bar{\xi})^{-1} \cL_Z \right)\partial_{\bar{\xi}} \hat{g} \hat{g}^{-1}- \left( \partial_Z \cL_{\bar{\xi}}-(\xi-\bar{\xi})^{-1} \cL_Z  \right)\partial_\xi \hat{g} \hat{g}^{-1}\right] .
\end{aligned}
\end{equation}
The appearance of $\cL_{Z}$ in this expression may be surprising as it is related to the $W$-component which we saw decouples from the theory. While is it valid to fix this component to zero from the beginning, this is unnecessary as it drops out on its own. This can be seen as follows. In order to use our solutions for the $\bCP^1$-dependence of $\cL$, we need to change basis using 
\begin{align*}
    (  \partial_Z \cL_\xi  +(\xi-\bar{\xi})^{-1} \cL_Z ) \big\vert_{Z=0}  &=\partial_Z \mathring{\cL}_\xi \big\vert_{Z=0} \\
    ( \partial_Z \cL_{\bar{\xi}}-(\xi-\bar{\xi})^{-1} \cL_Z ) \big\vert_{Z=0}  &=\partial_Z \mathring{\cL}_{\bar{\xi}} \big\vert_{Z=0} .
\end{align*}
Substituting \eqref{eq:L_of_UV} and \eqref{eq:UV_sol} and integrating over $\bCP^1$ with the help of the delta functions, we arrive at
\begin{align}
    \frac{1}{2} \int_{\mathbb{R}^2}  \dr \xi \wedge \dr \bar{\xi} ~(\xi-\bar{\xi}) ~ \tr ( \partial_\xi g g^{-1}-\partial_\xi \tilde{g} \tilde{g}^{-1}) ( \partial_{\bar{\xi}} g g^{-1}-\partial_{\bar{\xi}} \tilde{g} \tilde{g}^{-1}).
\end{align}
This is the action of the 2dIFT and can be rewritten as 
\begin{equation}\label{eq:2d_IFT}
    S_{\text{2dIFT}} = -\frac{1}{2}\int_{\bR^{2}} \dr \rho \wedge \dr z ~ \rho ~ \tr[(G^{-1} \partial_\rho G)^2+(G^{-1} \partial_z G)^2] \ . 
\end{equation}
In particular, when the original 4dCS gauge field is valued in the Lie Algebra $\mathfrak{sl}(2,\bR)$, this is the $\sigma$-model derived from 4d stationary and axisymmetric GR in \eqref{2d_sigma}. It takes the form of a symmetric-space $\sigma$-model with a spacetime-dependent coupling given by $\rho$ in these coordinates. 

Having derived the 2dIFT describing stationary, axisymmetric gravity from 4dCS theory, one might ask what this buys you. First and foremost, many technical aspects of this integrable system are now encoded in terms the geometry of the underlying 4-manifold, and the Lax connection has a natural home as the fundamental gauge field of 4dCS theory. 
In terms of applications, the inverse scattering method is a natural place to start. It relies heavily on the Lax formalism and after nearly half a century remains one of the most powerful tools to find and classify exact solutions, particularly axially symmetric black hole geometries in $D=4$ and $D=5$. 
Solitonic solutions like these are a fairly generic feature of integrable models.
A new solution is generated from a ``seed'' solution by applying a $\bCP^{1}$-dependent gauge transformation to the Lax. In particular, the gauge transformation parameter does not have generic dependence on $\bCP^{1}$, but is a rational function with simple poles and constraints on the residues of these poles.

Given that the 4dCS description is adapted to the Lax formalism, one might hope that the inverse scattering method has a natural home in this theory. Indeed, there is a reasonable candidate for the origin of these transformations in the residual symmetries of our theory. Recall that our derivation of the 2dIFT involved imposing the gauge fixing condition $\cL_{\bar{Z}} = 0$ in equation \eqref{eq:int_gauge}. This imposes a substantial constraint on the internal gauge symmetries, namely 
\begin{equation}\label{eq:residual_internal}
    \check{h}^{-1} \pd_{\bar{Z}} \check{h} = 0 \ . 
\end{equation}
The solution to this constraint that is most commonly encountered in the literature on 4dCS theory is to take $\check{h}$ to be independent of $\bCP^{1}$. However, if we allow $\check{h}$ to be a rational function of $\bCP^{1}$ with singularities, there may be a wider class of residual symmetries solving this constraint. 
Generically such a transformation would alter the $\bCP^{1}$-dependence of the Lax connection, producing poles at unwanted locations, but appropriate conditions on the residues of $\check{h}$ should ensure that no problems arise \cite{Harnad:1983pv}. 
In fact, this discussion should apply more generally to all 4dCS theories and their associated 2dIFTs. 
We leave this interesting application of 4dCS theory as an avenue for future work.

\section{Origin of 2dIFT from 4dWZW} \label{sec:4dWZW}

\begin{equation*}
\begin{tikzpicture}
\node at (-2,1) {\textcolor{tab10lightgray}{\textbf{6dCS}}};
\node at (-2,-1) {\textbf{4dCS}};
\node at (2,1) {\textbf{4dIFT}};
\node at (2,-1) {\textbf{2dIFT}};
\draw[->, draw=tab10lightgray, very thick] (-1,1) -- (1,1);
\draw[->, very thick] (-1,-1) -- (1,-1);
\draw[->, draw=tab10lightgray, very thick, decorate, decoration={snake, segment length=12.5pt, amplitude=2pt}] (-2,0.75) -- (-2,-0.75);
\draw[->, draw=tab10red, very thick, decorate, decoration={snake, segment length=12.5pt, amplitude=2pt}] (2,0.75) -- (2,-0.75);
\end{tikzpicture}
\end{equation*}

In the previous section, we constructed and studied a 4dCS theory which describes stationary, axisymmetric gravity. 
The main strength of this formalism is that it makes the integrable structure of the 2d model manifest, allowing these aspects of the theory to be studied directly. 
Key details of this construction appear mysterious, including a spacetime-dependent branch cut in the spectral plane and an equivariance condition on the Lax connection. 
The goal of the following sections is to provide a geometric origin for each feature of the 4dCS theory. 
Our approach will be to realise the 2dIFT \eqref{eq:2d_IFT} as a reduction of an integrable 4d theory known to arise\footnote{For a review of the relationship between the 4dWZW model and 6dCS theory, see the appendix.} from a 6d Chern-Simons (6dCS) theory \cite{Costello:2020, Bittleston:2020hfv}. Having understood this spacetime reduction (represented in red in the diagram above), we can find the corresponding reduction on twistor space, the underlying manifold of 6dCS theory. 
Applying the twistor space reduction to 6dCS theory we recover the 4dCS setup described in the previous section. 

Beginning this journey, the integrable 4d theory in question is known as the 4d Wess-Zumino-Witten (4dWZW) model \cite{Donaldson:1985zz, Losev:1995cr} and is defined by the action 
\begin{equation}
    S_{\text{4dWZW}} = \frac{1}{2} \int_{\bR^{4}} \tr \big( G^{-1} \dr G \wedge {\star} G^{-1} \dr G \big) 
    - \int_{\bR^{4}} \mu \wedge \text{WZ}[G] \ , 
\end{equation}
where, for our purposes, $G$ is an $\text{SL}(2,\bR)$-valued field. Its equation of motion is $\mu \wedge \pd ( \bar{\pd} G G^{-1} ) = 0$. 
In these expressions, we have introduced a 2-form $\mu = \dr u^{1} \wedge \dr \bar{u}^{1} + \dr u^{2} \wedge \dr \bar{u}^{2}$ which is proportional to the K{\"a}hler form on $\bR^{4}$ equipped with the Euclidean metric. 
The presence of this 2-form breaks some of the spacetime symmetry of the theory: the action is only invariant under diffeomorphisms which preserve the K{\"a}hler form. 
This includes all translations on $\bR^{4}$, but only a subgroup $\text{U}(2) \subset \text{SO}(4)$ of rotations. 
This subgroup is generated by 
\begin{equation}
\begin{aligned}
    R_{0} = & \frac{\rmi}{2} \big( u^{1} \pd_{u^{1}} - \bar{u}^{1} \pd_{\bar{u}^{1}} + u^{2} \pd_{u^{2}} - \bar{u}^{2} \pd_{\bar{u}^{2}} \big) \ , \\ 
    R_{1} = & \frac{\rmi}{2} \big( u^{1} \pd_{u^{1}} - \bar{u}^{1} \pd_{\bar{u}^{1}} - u^{2} \pd_{u^{2}} + \bar{u}^{2} \pd_{\bar{u}^{2}} \big) \ , \\ 
    R_{2} = & \frac{\rmi}{2} \big( u^{2} \pd_{u^{1}} - \bar{u}^{2} \pd_{\bar{u}^{1}} + u^{1} \pd_{u^{2}} - \bar{u}^{1} \pd_{\bar{u}^{2}} \big) \ , \\ 
    R_{3} = & \frac{1}{2} \big( u^{2} \pd_{u^{1}} + \bar{u}^{2} \pd_{\bar{u}^{1}} - u^{1} \pd_{u^{2}} - \bar{u}^{1} \pd_{\bar{u}^{2}} \big) \ . 
\end{aligned}
\end{equation}
The central U$(1)$ is generated by $R_{0}$ while the other three generators form an SU$(2)$ subalgebra, $[R_{i} , R_{j}] = \varepsilon_{ijk} R_{k}$.

We would like to perform a dimensional reduction to two-dimensions, by the spacetime vector fields 
\begin{equation}\label{eq:Xphi_Xtau}
    X_{\phi} = R_{0} + R_{1} = \rmi \big( u^{1} \pd_{u^{1}} - \bar{u}^{1} \pd_{\bar{u}^{1}} \big) \ , \qquad 
    X_{\tau} = \rmi \big( \pd_{u^{2}} - \pd_{\bar{u}^{2}} \big) \ . 
\end{equation}
These vector fields are simpler in cylindrical coordinates, defined by $u^{1} = \rho e^{\rmi \phi}$ and $u^{2} = z + \rmi \tau$, in which they read
\begin{equation}
    X_{\phi} = \pd_{\phi} \ , \qquad 
    X_{\tau} = \pd_{\tau} \ . 
\end{equation}
Whilst it is not surprising that the requisite reduction vectors are identical to those used to obtain the 2dIFT from 4d GR, it does not seem to the authors that this had to be the case. These theories are defined on different manifolds: the spacetime of the WZW model is $\bR^4$ equipped with a flat metric, rather than a dynamical spacetime metric in the case of 4d GR. Nevertheless, in both cases the reduction requires a  restriction to stationary and axisymmetric solutions, which for the 4dWZW model imposes
\begin{equation}\label{eq:inv_cond}
      \pd_{\phi} G = 0 \ , \qquad 
    \pd_{\tau} G = 0 \ .
\end{equation}
This constraint on the group-valued field $G$ implies that the 2-form WZ[$G$] does not have support on either $\dr \phi$ or $\dr \tau$. In cylindrical coordinates, the K{\"a}hler form reads 
\begin{equation}
    \mu = 2\rmi (\rho ~ \dr \phi \wedge \dr \rho+ \dr \tau \wedge \dr z),
\end{equation}
so it cannot compensate by saturating both $\dr \phi$ and $\dr \tau$. Another way of saying this \cite{Costello:2021} is that the 2-torus parameterised by a compactification of $\phi$ and $\tau$ has zero K{\"a}hler volume, meaning that $\int_{\phi} \int_{\tau} \mu = 0$. As a result the WZ term does not contribute to the reduced 2d model. Imposing \eqref{eq:inv_cond} on the surviving term in $S_\text{4dWZW}$ and performing the reduction (contracting by $X_\phi$ and $X_\tau$), we obtain the now familiar two-dimensional action
\begin{equation}
    S_{\text{2dIFT}} = - \frac{1}{2}\int_{\bR^{2}} \dr \rho \wedge \dr z ~ \rho ~ \tr[(G^{-1} \partial_\rho G)^2+(G^{-1} \partial_z G)^2] \ .
\end{equation}
Whilst this is identical to the action we are looking for, there is a key difference between this theory and the theory which describes gravity. 
In the context of stationary axisymmetric GR, the matrix $G$ parameterises the metric components along the isometry directions. 
In particular, this means that $G$ should be symmetric, taking values in the coset $\text{SL}(2,\bR) / \text{SO}(2)$ rather than the whole of $\text{SL}(2,\bR)$. 
We can pick out the subgroup $\text{SO}(2) \subset \text{SL}(2,\bR)$ by defining a $\bZ_{2}$-automorphism of the algebra $\fg = \mathfrak{sl}(2,\bR)$ acting as 
\begin{equation}
    \eta : \fg \to \fg \ , \qquad \eta : x \mapsto -x^{T} \ . 
\end{equation}
Since this map squares to the identity, the eigenvalues of $\eta$ are $\pm 1$. 
We can decompose the algebra into the corresponding subspaces as 
\begin{equation}
    \fg = \fg_{0} \oplus \fg_{1} \ , \qquad \fg_{0} \cong \mathfrak{so}(2) \ . 
\end{equation}
Here $\fg_{0}$ denotes the subspace preserved by $\eta$, which coincides with $\mathfrak{so}(2)$. 
We would like to restrict our group element to live in the coset $\text{SL}(2,\bR) / \text{SO}(2)$. 
This would mean that the algebra element generating $G$ lives in $\fg_{1}$, the $-1$-eigenspace of $\eta$, which exponentiates to\footnote{With a slight abuse of notation, we are denoting the group automorphism with the same symbol as the algebra automorphism.} the constraint $\eta (G) = G^{-1}$. For the automorphism specified above, this translates to $(G^{-1})^T = G^{-1}$ implying that $G$ is symmetric as desired. 

One might try to impose this constraint by demanding that the action is invariant under the transformation $G \mapsto \eta (G) = G^{-1}$. 
However, we see that the cubic term proportional to $\mu$ is not invariant under this transformation, it picks up a minus sign. 
To compensate for this, we will simultaneously apply a discrete spacetime transformation which preserves $\mu$ up to a sign. 
Such a transformation which is compatible with our reduction vectors is
\begin{equation}
    \sigma : (\rho, \phi, z, \tau) \mapsto (\rho, -\phi, z, -\tau) \ . 
\end{equation}
This reflection generates a $\bZ_{2}$-action on spacetime. The action $S_{\text{4dWZW}}$ is invariant under the combination of $\sigma$ and $\eta$ provided that $\eta (G) = G^{-1}$. We demand that the group element obeys this constraint as part of our reduction, which restricts the resulting 2d field to the coset $\text{SL}(2,\bR) / \text{SO}(2)$. 

The 4dWZW model is integrable, and can be described by a 6dCS theory on twistor space (for an introduction to this model and the details of this relationship see the appendix). 
Having established a reduction of the 4dWZW model which lands on the 2dIFT for stationary and axisymmetric GR, we will use this as a bridge to connect to 6dCS theory. 
In fact, this was the method by which we derived the 4dCS model presented earlier.

\section{Lift of reduction vectors to twistor space} \label{sec:twistorreduction}

In this section, we lift the spacetime reduction by $X_\phi$, $X_\tau$ to a reduction of twistor space so that it can be applied to 6dCS theory in section \ref{sec:6dCS_to_4dCS}. In preparation for this endeavour, let us review some relevant aspects of the geometry of (Euclidean) twistor space, which we will denote by $\bPT$. This section closely follows the exposition in \cite{Mason:1996}.

As a real manifold, there is an isomorphism 
\begin{equation}
    \bPT \cong \bCP^{1} \times \bR^{4} \ . 
\end{equation}
On the Riemann sphere factor, we will use a complex coordinate $\zeta \in \bCP^{1}$, and we will use standard Cartesian coordinates $\{ x^{1} , x^{2} , x^{3} , x^{4} \}$ on $\bR^{4}$. It is also helpful to employ the complex combinations $u^{1} = x^{1} + \rmi x^{2}$ and $u^{2} = x^{3} + \rmi x^{4}$ which appeared in the previous section. 
While this is an isomorphism of real manifolds, meaning that these serve as good coordinates on twistor space, the complex structure on twistor space mixes the $\bCP^{1}$ and $\bR^{4}$ factors. Holomorphic coordinates on twistor space are given by 
\begin{equation}
    \zeta \ , \qquad 
    v^{1} = u^{1} - \zeta \, \bar{u}^{2} \ , \qquad 
    v^{2} = u^{2} + \zeta \, \bar{u}^{1} \ . 
\end{equation}
It is important to highlight that $\{ u^{1} , u^{2} \}$ are \textit{not} holomorphic coordinates on twistor space, though they coincide with $\{ v^{1} , v^{2} \}$ on the $\bR^{4}$ defined by $\zeta = 0$. Holomorphic coordinates on the northern patch are related to those on the southern patch by 
\begin{equation}
    \tilde{\zeta} = 1 / \zeta \ , \qquad 
    \tilde{v}^{1} = v^{1} / \zeta \ , \qquad 
    \tilde{v}^{2} = v^{2} / \zeta \ . 
\end{equation}

The relationship between twistor space and spacetime, known as the twistor correspondence, is most easily seen by considering the product manifold $\bPS = \bCP^{1} \times \bR^{4}$ known as the projective spin bundle. 
It is captured by the double fibration 
\begin{equation*}
\begin{tikzpicture}
\node at (0,1.5) {$\bPS$};
\node at (-2,0) {$\bPT$};
\node at (2,0) {$\bR^{4}$};
\draw[->] (-0.35,1.25) -- (-1.65,0.25);
\node at (-1.2,0.9) {$p$};
\draw[->] (0.35,1.25) -- (1.65,0.25);
\node at (1.2,0.9) {$q$};
\end{tikzpicture}
\end{equation*}

\noindent
where the maps $p$ and $q$ are given by 
\begin{equation}
    p : (\zeta, x^i) \mapsto (\zeta, v^{1}, v^{2}) \ , \qquad 
    q : (\zeta, x^i) \mapsto x^i \ . 
\end{equation}
We should highlight that this correspondence space and double fibration exist for all signatures of metric on $\bR^{4}$. In fact, it is best understood by working with a complex metric on $\bC^{4}$ which becomes $\bR^{4}$ with various signatures when restricted to certain real slices. 
The isomorphism between twistor space $\bPT$ and the projective spin bundle $\bPS$ does not hold for general signatures, however. This isomorphism only exists in Euclidean signature when $p : \bPS \to \bPT$ is invertible. It is often convenient to work in Euclidean signature, which is the approach we take here, and then analytically continue the result to other signatures of interest.

Since we are interested in performing a reduction of twistor space, we need to know how various conformal transformations on $\bR^{4}$ lift to transformations of twistor space. In fact, the strategy which we will adopt is to first lift the conformal Killing vectors to the correspondence space $\bPS$, and then to project these vectors down to twistor space using the map $p : \bPS \to \bPT$. A general spacetime vector field may be expressed as 
\begin{equation}
    X = X^{u^{1}} \pd_{u^{1}} + X^{\bar{u}^{1}} \pd_{\bar{u}^{1}} + X^{u^{2}} \pd_{u^{2}} + X^{\bar{u}^{2}} \pd_{\bar{u}^{2}} \ . 
\end{equation}
We will assume that this is a conformal Killing vector for the Euclidean spacetime metric $\dr s^{2} = \dr u^{1} \, \dr \bar{u}^{1} + \dr u^{2} \, \dr \bar{u}^{2}$. If this is to be a real vector field on $\bR^{4}$, then these components must also obey the relations $X^{\bar{u}^{a}} = \overline{X^{u^{a}}}$. 
We would like to lift this to a vector field $X^{\prime \prime}$ on $\bPS$ which satisfies $q_{\ast} (X^{\prime \prime}) = X$. We will choose this lift such that the projection $X^{\prime} = p_{\ast} (X^{\prime \prime})$ is a well-defined holomorphic vector field on twistor space. 
\begin{equation*}
\begin{tikzpicture}
\node at (0,1.5) {$X^{\prime \prime}$};
\node at (-2,0) {$X^{\prime}$};
\node at (2,0) {$X$};
\draw[->] (-0.35,1.25) -- (-1.65,0.25);
\node at (-1.2,0.9) {$p_{\ast}$};
\draw[->] (0.35,1.25) -- (1.65,0.25);
\node at (1.2,0.9) {$q_{\ast}$};
\end{tikzpicture}
\end{equation*}

Vector fields on $\bPS$ which are tangent to the projection map $p : \bPS \to \bPT$ are given by 
\begin{equation}
    V_{1} = \pd_{\bar{u}^{1}} - \zeta \, \pd_{u^{2}} \ , \qquad 
    V_{2} = \pd_{\bar{u}^{2}} + \zeta \, \pd_{u^{1}} \ . 
\end{equation}
For each fixed $\zeta \in \bCP^{1}$, one can show (using the conformal Killing equation) that 
\begin{equation}
    [ X , V_{1} ] = Q \, \pd_{u^{2}} \ , \qquad 
    [ X , V_{2} ] = -Q \, \pd_{u^{1}} \ , \qquad 
    \mod \{ V_{1}, V_{2} \} 
\end{equation}
where $Q$ is given by 
\begin{equation}
    Q = \pd_{\bar{u}^{2}} X^{u^{1}} + \zeta \big( \pd_{u^{1}} X^{u^{1}} - \pd_{\bar{u}^{2}} X^{\bar{u}^{2}} \big) - \zeta^{2} \pd_{u^{1}} X^{\bar{u}^{2}} \ . 
\end{equation}
The conformal Killing equation also shows that $Q$ is constant along $V_{1}$ and $V_{2}$, meaning that the lifted vector field 
\begin{equation}
    X^{\prime \prime} = X + Q \, \pd_{\zeta} + \bar{Q} \, \pd_{\bar{\zeta}} \ , 
\end{equation}
satisfies $[ X^{\prime \prime} , V_{1} ] = 0$ and $[ X^{\prime \prime} , V_{2} ] = 0$ modulo $\{ V_{1}, V_{2} , \pd_{\bar{\zeta}} \}$.  The component involving $\bar{Q}$ ensures that this is a real vector field. The projection of this vector field to twistor space is explicitly given by 
\begin{equation}
    X^{\prime} = \big( X^{u^{1}} - \zeta \, X^{\bar{u}^{2}} - \bar{u}^{2} \, Q \big) \pd_{v^{1}} 
    + \big( X^{u^{2}} + \zeta \, X^{\bar{u}^{1}} + \bar{u}^{1} \, Q \big) \pd_{v^{2}} + Q \, \pd_{\zeta} \ , 
\end{equation}
where the components are constant along $V_{1}$ and $V_{2}$ and hence holomorphic functions of $\{ \zeta, v^{1}, v^{2} \}$. 
The lifts and projections of some conformal Killing vectors which generate the symmetries of the 4dWZW model are presented in table \ref{tab:ConformalVectors}. 
\begin{table}[ht]
\centering
\begin{equation*}
\begin{array}{c|c|c}
    X & Q & X^{\prime} \\ 
    \hline
    \pd_{u^{1}} & - & \pd_{v^{1}} \\ 
    \pd_{\bar{u}^{1}} & - & \zeta \pd_{v^{2}} \\ 
    \pd_{u^{2}} & - & \pd_{v^{2}} \\ 
    \pd_{\bar{u}^{2}} & - & -\zeta \pd_{v^{1}} \\ 
    R_{0} & \rmi \zeta & (\rmi/ 2) ( v^{1} \pd_{v^{1}} + v^{2} \pd_{v^{2}} ) + \rmi \zeta \pd_{\zeta} \\ 
    R_{1} & - & (\rmi/ 2) ( v^{1} \pd_{v^{1}} - v^{2} \pd_{v^{2}} ) \\ 
    R_{2} & - & (\rmi/ 2) ( v^{2} \pd_{v^{1}} + v^{1} \pd_{v^{2}} ) \\ 
    R_{3} & - & (1/2) ( v^{2} \pd_{v^{1}} - v^{1} \pd_{v^{2}} )
\end{array}
\end{equation*}
\caption{Lifts and projections of some conformal Killing vectors.}
\label{tab:ConformalVectors}
\end{table}

Having understood how to move vector fields between these spaces, we would like to study the twistor correspondence under a pair of reductions.
As a warm up, let us consider a reduction by two translational isometries, given by the spacetime vector fields $X = \pd_{u^{2}}$ and $Y = \pd_{\bar{u}^{2}}$. On the $\bR^{4}$, the quotient space is identified with $\bR^{2}$ and parameterised by coordinates $\{ u^{1} , \bar{u}^{1} \}$. The lift of these vector fields to the correspondence space $\bPS =\bCP^{1} \times \bR^{4}$ is trivial, so the reduced correspondence space is $M_{4} = \bCP^{1} \times \bR^{2}$. Finally, the projection of these vector fields to twistor space gives $X^{\prime} = \pd_{v^{2}}$ and $Y^{\prime} = -\zeta \pd_{v^{1}}$. Functions on the quotient space must be independent of $\{ v^{1} , v^{2} \}$, and so the remaining coordinate $\zeta$ is a holomorphic coordinate on the reduced twistor space, which is therefore identified with $\bCP^{1}$. In summary, we have a reduced twistor correspondence captured by the following diagram. 
\begin{equation*}
\begin{tikzpicture}
\node at (0,1.5) {$\bPS$};
\node at (0,-1) {$M_{4}$};
\node at (-2,0) {$\bPT$};
\node at (2,0) {$\bR^{4}$};
\node at (-2,-2.5) {$\bCP^{1}$};
\node at (2,-2.5) {$\bR^{2}$};
\draw[->] (-0.35,1.25) -- (-1.65,0.25);
\draw[->] (0.35,1.25) -- (1.65,0.25);
\draw[->] (-0.35,-1.25) -- (-1.65,-2.25);
\draw[->] (0.35,-1.25) -- (1.65,-2.25);
\draw[->, decorate, decoration={snake, segment length=8.8pt, amplitude=1.5pt}] (0,1.25) -- (0,-0.75);
\draw[->, decorate, decoration={snake, segment length=8.8pt, amplitude=1.5pt}] (-2,-0.25) -- (-2,-2.25);
\draw[->, decorate, decoration={snake, segment length=8.8pt, amplitude=1.5pt}] (2,-0.25) -- (2,-2.25);
\end{tikzpicture}
\end{equation*}
Now, let us move to the reduction of interest, which is generated by the spacetime vector fields $X_{\phi}$, $X_{\tau}$ introduced in \eqref{eq:Xphi_Xtau}. It is once more convenient to work in cylindrical coordinates (defined by $u^{1} = \rho e^{\rmi \phi}$ and $u^{2} = z + \rmi \tau$) in which these vector fields are $X_{\phi} = \pd_{\phi}$ and $X_{\tau} = \pd_{\tau}$. Coordinates on the quotient space of $\bR^{4}$ are provided by $\{ \rho , z \}$, and it is identified with $\bR_{\geq 0} \times \bR$. The lift of the vector fields to the correspondence space is given by 
\begin{equation}
    X_{\phi}^{\prime \prime} = \pd_{\phi} + \rmi \zeta \pd_{\zeta} - \rmi \bar{\zeta} \pd_{\bar{\zeta}} \ , \qquad 
    X_{\tau}^{\prime \prime} = \pd_{\tau} \ . 
\end{equation}
Unlike for the translational isometries just considered, for the rotational isometry we get a non-trivial mixing of the $\bCP^1$ and $\bR^4$ coordinates, i.e. $Q= \rmi \zeta$. 
We would like to find coordinates on the quotient space $M_{4}$ which are invariant under the flow of these vector fields. The obvious candidate for the $\bCP^{1}$ factor, $\zeta$, is not invariant, but by mixing the $\bCP^{1}$ and $\bR^{4}$ coordinates as $Z = e^{-\rmi \phi} \zeta$ we obtain an invariant spectral parameter. We can continue to use coordinates $\{ \rho , z \}$ on the spacetime factor in $M_{4}$. 

The projection of these vector fields to twistor space is then
\begin{equation}
    X_{\phi}^{\prime} = \rmi ( v^{1} \pd_{v^{1}} + \zeta \pd_{\zeta} ) \ , \qquad 
    X_{\tau}^{\prime} = \rmi ( \pd_{v^{2}} - \zeta \pd_{v^{1}} ) \ . 
\end{equation}
The reduced twistor space should be a one-dimensional complex manifold with a single holomorphic coordinate. Indeed the quotient of $\bPT$ by $X_\phi'$, $X_\tau'$ may be parameterised by 
\begin{equation}
    W = \frac{1}{2} \bigg( v^{2} + \frac{v^{1}}{\zeta} \bigg) = z + \frac{\rho}{2} \big( Z^{-1} - Z \big) \ , 
\end{equation}
where the first expression shows that this coordinate is holomorphic, and its invariance can be easily verified by acting with the reduction vectors. Meanwhile, the second expression exhibits the relationship of this spectral parameter to the invariant spectral parameter $Z$. We recognise this relationship as the two-to-one covering map from our 4dCS construction. It reveals that the $\bCP^1$ coordinatized by $Z$ is that of the reduced projective spin bundle, while the $\bCP^1$ coordinatized by $W$ is that of the reduced twistor space. 
The explicit relationship between these variables was derived by imposing invariance under the reduction vectors, and follows from our particular choice of reduction as well as the twistor correspondence. 
Alternative reductions would lead to different relationships between these parameters. 

It remains to identify the lift of the discrete transformation 
\begin{equation}
    \sigma : (\rho, \phi, z, \tau) \mapsto (\rho, -\phi, z, -\tau) \ . 
\end{equation}
We will follow the same rational as above, lifting this to a transformation of $\bPS$ subject to the constraint that it projects down to a holomorphic map on $\bPT$. In the complex coordinates on $\bR^{4}$, this discrete transformation acts as $\sigma : (u^{1}, u^{2}) \mapsto (\bar{u}^{1} , \bar{u}^{2})$. 
Inspecting the expressions $v^{1} = u^{1} - \zeta \, \bar{u}^{2}$ and $v^{2} = u^{2} + \zeta \, \bar{u}^{1}$, we see that $\sigma$ exchanges these combinations (up to an overall rescaling) if we include the map $\zeta \mapsto - \zeta^{-1}$. We therefore define the lift of this discrete transformation as
\begin{equation}
    \sigma^{\prime \prime} : (\zeta , \rho, \phi, z, \tau) \mapsto (- \zeta^{-1} , \rho, -\phi, z, -\tau) \ . 
\end{equation}
The action on the invariant spectral parameter $\sigma^{\prime \prime} : Z \mapsto - Z^{-1}$ moves between the two sheets of the double covering over the $W$-plane. The projection down to twistor space acts as 
\begin{equation}
    \sigma^{\prime} : (\zeta , v^{1} , v^{2}) \mapsto \bigg( -\frac{1}{\zeta} , -\frac{v^{2}}{\zeta} , \frac{v^{1}}{\zeta} \bigg) \ . 
\end{equation}
This is a holomorphic map, as required, and it acts trivially on the spectral parameter $W$.

\section{Reduction of 6dCS theory}\label{sec:6dCS_to_4dCS}

\begin{equation*}
\begin{tikzpicture}
\node at (-2,1) {\textbf{6dCS}};
\node at (-2,-1) {\textbf{4dCS}};
\node at (2,1) {\textbf{4dIFT}};
\node at (2,-1) {\textbf{2dIFT}};
\draw[->, very thick] (-1,1) -- (1,1);
\draw[->, very thick] (-1,-1) -- (1,-1);
\draw[->, draw=tab10red, very thick, decorate, decoration={snake, segment length=12.5pt, amplitude=2pt}] (-2,0.75) -- (-2,-0.75);
\draw[->, very thick, decorate, decoration={snake, segment length=12.5pt, amplitude=2pt}] (2,0.75) -- (2,-0.75);
\end{tikzpicture}
\end{equation*}

In the previous section, we studied geometric aspects of the reduction. We lifted the spacetime reduction vectors to the correspondence space $\bPS$ and then projected them down to holomorphic vectors on twistor space $\bPT$. Applying this reduction to each of these manifolds, we exhibited a reduced twistor correspondence with correspondence space $M_{4} = \bCP^{1} \times \bR^{2}$. This is the 4-manifold over which our 4dCS theory is defined, and in this section we will recover that theory as a reduction of 6dCS theory.

Six-dimensional Chern-Simons (6dCS) theory \cite{Costello:2020, Bittleston:2020hfv, Penna:2020uky, Costello:2021} is to 4d integrable theories as 4dCS theory is to 2d integrable theories. 
The underlying 6-manifold is isomorphic to $\bCP^{1} \times \bR^{4}$ and there is a localisation procedure by which one may integrate out the $\bCP^{1}$. In particular, the 4dWZW model presented in section \ref{sec:4dWZW} is known to admit a 6dCS description \cite{Costello:2020, Bittleston:2020hfv}.
A pedagogical introduction to this theory, and the localisation to the 4dWZW model are presented in the appendix. Since those results are not novel to this work, we will skip the majority of the details here and jump directly to the reduction to 4dCS theory. The action of 6dCS theory is given by 
\begin{equation}
    S_{\text{6dCS}} = \frac{1}{2 \pi \rmi} \int_{\bPT} \Omega \wedge \tr \bigg( \cA \wedge \bar{\pd} \cA + \frac{2}{3} \cA \wedge \cA \wedge \cA \bigg) \ , 
\end{equation}
where the meromorphic 3-form $\Omega$ and boundary conditions are given by 
\begin{equation}
    \Omega = \frac{\dr \zeta \wedge \dr v^{1} \wedge \dr v^{2}}{\zeta^{2}} \ , \qquad
    \cA \big\vert_{\zeta = 0} = 0 \ , \qquad
    \cA \big\vert_{\tilde{\zeta} = 0} = 0 \ .
\end{equation}
Much like in 4dCS theory, the choice of meromorphic form and boundary conditions are input data, and different choices lead to different integrable field theories --- now four-dimensional theories (4dIFTs) rather than two-dimensional. The fundamental gauge field is an algebra-valued $(0,1)$-form, where we are making use of the complex structure on twistor space\footnote{Since the reduction from 6dCS to 4dCS takes place on the correspondence spaces, we think it is more precise to define 6dCS theory over $\bPS$. In this case, the action would be defined by first pulling $\Omega$ back via $p : \bPS \to \bPT$ and then wedging against the Chern-Simons 3-form. The gauge field and exterior derivative would initially be considered as generic, and then one observes that the legs along $\Omega$ drop out of the action and decouple from the theory. This is corroborated by the fact that $\bPT$ changes dimension depending on the signature of the metric on $\bR^{4}$, an issue which was circumvented in \cite{Bittleston:2020hfv} by moving to the correspondence space. We argue that this is the correct approach, even for Euclidean signature (where $\bPS$ and $\bPT$ are isomorphic as real manifolds).} to introduce the Dobeault complex. In this notation, a $(p,q)$-form is a $(p+q)$-form with $p$ legs along the holomorphic directions and $q$ legs along the anti-holomorphic directions. Similarly, the Dobeault operator in the action is defined by 
\begin{equation}
    \bar{\pd} : \Omega^{p,q} \to \Omega^{p,q+1} \ , \qquad \bar{\pd} = \pi_{p,q+1} \circ \dr \ , 
\end{equation}
where $\pi_{p,q}$ is the projection from the space of $(p+q)$-forms to the subspace of $(p,q)$-forms. The 4dIFT associated to this theory can be derived from a similar localisation procedure to that of section \ref{sec:4dCS}, performing the integral over $\bCP^{1}$ explicitly after solving the bulk equations of motion. As already noted, a more generous introduction to 6dCS theory and the details of this localisation are given in the appendix. 

In the remainder of this section, we seek to close the commutative diagram above by reducing 6dCS theory on the lifted killing vectors described in section \ref{sec:twistorreduction}. Recall that these can be expressed as 
\begin{equation}
    X_{\phi}^{\prime \prime} = \pd_{\phi} + \rmi \zeta \pd_{\zeta} - \rmi \bar{\zeta} \pd_{\bar{\zeta}} \ , \qquad 
    X_{\tau}^{\prime \prime} = \pd_{\tau} \ . 
\end{equation}
Our reduction procedure follows the methodology presented in \cite{Bittleston:2020hfv}. First, we impose invariance of the gauge field under the action of these reduction vectors, that is 
\begin{equation}
    \cL_{X_{\phi}^{\prime \prime}} \cA = 0 \ , \qquad
    \cL_{X_{\tau}^{\prime \prime}} \cA = 0 \ . 
\end{equation}
Having done this, our Lagrangian 6-form will be invariant under the flow generated by these vector fields. We would like to compute the associated 4-form on the surviving quotient space. In practice, this is achieved by contracting the Lagrangian with the bivector $X_{\tau}^{\prime \prime} \wedge X_{\phi}^{\prime \prime}$. 

There are two technical preparations we can make in order to facilitate this computation. 
The complex structure on twistor space mixes the $\bCP^{1}$ and $\bR^{4}$ directions, meaning that the natural $(1,0)$-forms $\{\dr v^1, \dr v^2\}$ have legs along both directions. 
For our purposes, it will be helpful to implement the splitting between spacetime and the spectral parameter by introducing a basis of forms which only have legs along either $\bCP^{1}$ or $\bR^{4}$ but not both. 
We would still like these to provide a basis for the $(1,0)$-forms and $(0,1)$-forms with respect to the complex structure on twistor space, and it simplifies the computation further if they are invariant under $X_\phi''$ and $X_\tau''$. 
One finds that a suitable basis of $(1,0)$-forms is given by 
\begin{align}
    \eta^0=e^{-\rmi \phi} \dr \zeta \ , \qquad 
    \eta^1=e^{-\rmi \phi} (\dr u^1 - \zeta \, \dr \bar{u}^2) \ , \qquad 
    \eta^2=\dr u^2 + \zeta \, \dr \bar{u}^1 \ . 
\end{align}
Note that the functional dependence on $\zeta$ and $\phi$ can be repackaged into the parameter $Z = e^{-\rmi \phi} \zeta$ which is invariant under the flow of our reduction vectors. 
Similarly, for $(0,1)$-forms we can use
\begin{align}
    \bar{\eta}^0=e^{\rmi \phi} \dr \bar{\zeta} \ , \qquad \bar{\eta}^1= \frac{e^{\rmi \phi} (\dr \bar{u}^1 - \bar{\zeta} \, \dr u^2) }{1+\zeta \bar{\zeta}} \
    , \qquad \bar{\eta}^2= \frac{\dr \bar{u}^2 + \bar{\zeta} \, \dr u^1  }{1+\zeta \bar{\zeta}} \ .
\end{align}
In this basis, the components of the gauge field may be expressed as 
\begin{equation}
    \cA = \cA_{0} \, \bar{\eta}^0 + \cA_{1} \, \bar{\eta}^1 + \cA_{2} \, \bar{\eta}^2 \ . 
\end{equation}
The fact that the basis forms are invariant means that the constraints $\cL_{X_{\phi}^{\prime \prime}} \cA = 0$ and $\cL_{X_{\tau}^{\prime \prime}} \cA = 0$ amount to the individual components of $\cA$ being independent of the isometry coordinates $\{ \phi , \tau \}$.

In addition, it is helpful to ``prepare the gauge field for reduction'' by imposing the gauge fixing constraints 
\begin{equation}
    X_{\phi}^{\prime \prime} \vee \cA = 0 \ , \qquad
    X_{\tau}^{\prime \prime} \vee \cA = 0 \ . 
\end{equation}
This ensures that the bivector $X_{\tau}^{\prime \prime} \wedge X_{\phi}^{\prime \prime}$ contracting the Lagrangian acts only on $\Omega$, dramatically simplifying the calculation. 
This constraint may be realised using the $(1,0)$-form shift symmetry 
\begin{equation}
    \cA \mapsto \cA + \big( C_{0} \, \bar{\eta}^0 + C_{1} \, \bar{\eta}^1 + C_{2} \, \bar{\eta}^2 \big) \ . 
\end{equation}
This is a trivial symmetry of the action which leaves it invariant because $\Omega$ saturates the $(3,0)$-legs. Solving the constraints above for the components of the $(1,0)$-form $C$, one finds an expression for the shifted gauge field which is most easily expressed in terms of the variables $Z = e^{-\rmi \phi} \zeta$ and $\xi = z + \rmi \rho$. In these coordinates, the shifted gauge field may be written as 
\begin{equation}
    \cA + C = A_{\bar{Z}} \dr \bar{Z} + A_{\xi} \dr \xi + A_{\bar{\xi}} \dr \bar{\xi} \ . 
\end{equation}
The precise dependence of the new components $\{ A_{\bar{Z}} , A_{\xi} , A_{\bar{\xi}} \}$ on the old components is not especially important as these still represent generic field configurations of the theory. What is important, however, is the analytic behaviour of these components in $Z \in \bCP^{1}$. One finds that the boundary conditions on $\cA$ imply that $A_{\xi}$ and $A_{\bar{\xi}}$ vanish at both $Z = 0$ and $Z = \infty$. Furthermore one also observes that $A_{\xi}$ is permitted a simple pole at $Z = -\rmi$, whilst $A_{\bar{\xi}}$ is permitted a simple pole at $Z = +\rmi$. This matches the properties of the 4dCS gauge field presented in section \ref{sec:4dCS}.

After making these preparations, computing the contraction of the bivector $X_{\tau}^{\prime \prime} \wedge X_{\phi}^{\prime \prime}$ 
with the Lagrangian 6-form amounts to computing its contraction with $\Omega$. This yields the meromorphic 1-form of our 4dCS theory which is given by 
\begin{equation}
    \omega = \frac{1}{2} (X_{\tau}^{\prime \prime} \wedge X_{\phi}^{\prime \prime}) \vee \Omega = \dr \bigg( z + \frac{\rho}{2} \big( Z^{-1} - Z \big) \bigg) \ . 
\end{equation}
In summary, we have landed on the 4dCS theory described in section \ref{sec:4dCS} with the action 
\begin{equation}
    S_{\text{4dCS}} = \frac{1}{2 \pi \rmi} \int_{M_{4}} \omega \wedge \tr \bigg( A \wedge \dr A + \frac{2}{3} A \wedge A \wedge A \bigg) \ . 
\end{equation}

The final ingredient to consider is the discrete reduction generated by a combination of the reflection 
\begin{equation}
    \sigma^{\prime \prime} : (\zeta , \rho, \phi, z, \tau) \mapsto (- \zeta^{-1} , \rho, -\phi, z, -\tau) \ . 
\end{equation}
and the $\bZ_{2}$-automorphism $\eta : \fg \to \fg$. The action on the isometry coordinates $\phi$ and $\tau$ does not play a role in the 4dCS theory, but the reflection does act non-trivially on the invariant spectral parameter as $Z \mapsto -Z^{-1}$. Demanding that the 6dCS gauge field also be invariant under this discrete reduction implies that the 4dCS gauge field should satisfy 
\begin{equation}
    A(Z) = \eta \big( A(-Z^{-1}) \big) \ . 
\end{equation}
This is the equivariance condition which reduced the Lie group $\text{SL}(2,\bR)$ of the 2d theory to the coset $\text{SL}(2,\bR) / \text{SO}(2)$. In the context of 4dCS theory, these equivariance conditions are related to branch cut defects \cite{Costello:2019tri}. In this section, we have shown that branch cut defects in 4dCS theory arise from discrete reductions of 6dCS theory. Since this is interesting in its own right, we devote appendix \ref{app:symm_PCM} to presenting this result in the simpler case of the symmetric space PCM.

\section{Outlook}

The focus of this work has been on establishing a firm basis for the study of integrable sectors of gravity in terms of 4d Chern-Simons theory. While we have often referred explicitly to 4d vacuum GR with one spacelike and one timelike killing vector, our construction can be applied to the integrable sectors of many different gravity theories. This is due to the observation by Breitenlohner, Maison, and Gibbons \cite{Breitenlohner:1987dg} that 
with $D-2$ commuting Killing vectors, the bosonic sectors of various $D$-dimensional 
supergravities yield 2d $\sigma$-models of the same form. By taking the field $G$ to be valued in the relevant coset group, and the Chern-Simons fields in the associated Lie algebra, our analysis applies to each such case. 

To describe the full field content of a given supergravity theory the corresponding 2d $\sigma$-model must be coupled to fermions (see for example \cite{Nicolai:1987kz}). This is one motivation to incorporate fermionic degrees of freedom into the commutative diagram of models discussed in this paper.
One approach to this problem in 6dCS theory was presented in \cite{Penna:2020uky}, similar in spirit to the \textit{order defects} of 4dCS theory described in \cite{Costello:2019tri}. 
This introduced sufficient degrees of freedom in the 2dIFT to describe the reduction of 4d $N=1$ supergravity, up to quadratic order in the fermions. 
In general, the topic of order defects in 6dCS theory and their reduction to 4dCS theory requires further development.

Dressing transformations play a key role in generating new solutions in gravity, and have many applications in integrable models more generally where solitonic solutions are a common feature. 
As these transformations are adapted to the Lax formalism, it is natural to study them from the perspective of the associated 4dCS theories. 
As alluded to in section \ref{sec:4dCS}, we suspect that dressing transformations are related to residual gauge symmetries with simple poles in the spectral plane. 
One might wonder whether the constraints imposed in the BZ method and its higher-dimensional generalisations have a natural origin in the 4dCS model. 
For example, requiring that the analytic structure of the Lax connection is preserved implies that the residues of simple poles are rank 1 matrices \cite{Harnad:1983pv}. 
Based on \cite{Katsimpouri:2013wka}, extensions to more general coset models for supergravity theories will require poles with higher-rank residues. 
Ultimately, we hope the 4dCS perspective can shed light on solitonic solutions, both in 4d vacuum gravity and higher-dimensional supergravity theories.

Asymptotics are a topic which is extensively studied in the literature on inverse scattering methods but less understood in the 4dCS context. Hopefully large $(\rho,z)$ conditions which are desirable from a physical standpoint can be ensured by corresponding assumptions on the CS gauge field. There is the added subtlety that the 2d spacetime of interest is often the half-plane, rather than $\mathbb{R}^2$, and thus has a boundary. This may provide an interesting arena to study integrable boundary conditions in the context of 4dCS theory.

One of the advantages of the 4dCS framework is that it facilitates a systematic classification of seemingly unrelated integrable models in terms of the properties of the one-form $\omega$. In this work, we have introduced several new ingredients which can help expand that classification. First, we have shown how discrete reductions can be used alongside reductions by Killing vectors, not only to obtain 2dIFTs valued in symmetric spaces from 4dIFTs, but also to obtain the corresponding 4dCS theory from 6dCS. The discrete reduction has the effect of introducing a branch cut in the spectral plane. In appendix \ref{app:symm_PCM}, we isolate this ingredient and show how to recover branch cut defects in 4dCS theory in the simpler example of the symmetric space PCM.

Reduction on an angular coordinate introduces a second new ingredient to 4dCS theory: mixed $\bCP^1$ and spacetime dependence in the one-form $\omega$. It is the combination of reduction by an axial Killing vector and discrete reduction which leads to spacetime-dependent branch cuts in the $W$-plane, or mixing of $\bCP^1$ and spacetime derivatives with respect to its double cover $Z$. This is the significance (from the CS perspective) of the different formulations of the Lax for stationary axisymmetric GR in terms of a variable or constant spectral parameter, in \cite{Belinsky:1979mh} compared with \cite{Breitenlohner:1986um}, for example.

Both of these ingredients may be used in the future to construct new 4dCS models and their associated 2dIFTs. For example, recent work on 6dCS theory \cite{Cole:2023umd} showed how to recover the $\lambda$-deformation \cite{Sfetsos:2013wia} by splitting the double pole at $\zeta = 0$ into two simple poles. This would break the discrete symmetry $\zeta \mapsto - \zeta^{-1}$ unless one also splits the double pole at $\zeta = \infty$ into two simple poles. We suspect that applying a discrete reduction to the 6dCS setup with four simple poles may result in the symmetric-space $\lambda$-model whose 4dCS description has been explored in \cite{Tian:2020ryu, Schmidtt:2020dbf}. Gauged integrable field theories were also derived from 6dCS theory in \cite{Cole:2024ess}. It would be interesting to see if symmetric-space models can be recovered using that formalism, and how it relates to the methodology presented in this paper.

In addition, the admissibility of meromorphic 1-forms in 4dCS theory with spacetime dependence may be relevant to the 2dIFTs recently introduced in \cite{Hoare:2020fye}. In that work, a variety of integrable models were studied in which the couplings were allowed to depend on the 2d spacetime coordinates. It was shown that these models were classically integrable, provided that the couplings took a special form, related to the RG flow equations. Using the results of this paper, it may now be possible to incorporate these models into the 4dCS formalism. This could be particularly interesting in light of the recent work relating RG flow and 4dCS theory \cite{Delduc:2020vxy, Derryberry:2021rne, Levine:2023wvt, Lacroix:2024wrd}. These papers propose a mechanism for computing the RG flow of a given 2dIFT very directly, supposing that one knows the appropriate 4dCS description.

\section*{Acknowledgements}

The authors would like to thank Ibrahima Bah, Roland Bittleston, Wei Bu, Ryan Cullinan, Ben Hoare, Timothy Hollowood, Joaquin Liniado, James Lucietti, Jose Luis Miramontes, and Daniel Thompson for insightful discussions. The work of PW is supported by the grant ST/X000648/1.
 \vspace{4mm}
 
 \noindent \textbf{Open Access Statement}--- For the purpose of open access, the authors have applied a Creative Commons Attribution (CC BY) licence to any Author Accepted Manuscript version arising.

%% file: appendix.tex
\section{Pedagogical introduction to 6dCS theory}

\subsection{Penrose-Ward transform} \label{sec:PWtransform}

Integrable theories in four dimensions are described by the self-dual Yang-Mills (sdYM) equation 
\begin{equation}
    F = \star F \ , \qquad
    F = \dr A + \frac{1}{2} [A,A] \ . 
\end{equation}
Solutions to this equation on $\bR^{4}$ are related to twistor space via the Penrose-Ward transform which we will briefly review. To describe this relationship, we start by rewriting the sdYM equation as three independent equations, 
\begin{equation}
    \dr u^{1} \wedge \dr u^{2} \wedge F = 0 \ , \qquad 
    \dr \bar{u}^{1} \wedge \dr \bar{u}^{2} \wedge F = 0 \ , \qquad 
    \mu \wedge F = 0 \ . 
\end{equation}
We have introduced a 2-form $\mu = \dr u^{1} \wedge \dr \bar{u}^{1} + \dr u^{2} \wedge \dr \bar{u}^{2}$ which is proportional to the K{\"a}hler form on $\bR^{4}$. This rewriting of the sdYM equation exploits some pleasant properties of 2-forms in four dimensions. These three equations may be recast as a single equation by introducing an auxiliary complex parameter, 
\begin{equation} \label{eq:twistorsdYM}
    \dr \zeta \wedge \dr v^{1} \wedge \dr v^{2} \wedge F = 0 \qquad \forall \ \zeta \in \bCP^{1} \ . 
\end{equation}
Let us think of this as an equation on twistor space for a $\bCP^{1}$-independent gauge field which only has legs along $\bR^{4}$. In this spirit, we can decompose the $\bR^{4}$ gauge field into its $(1,0)$ and $(0,1)$ components on twistor space, 
\begin{equation}
\begin{aligned}
    A^{1,0} & = 
    \frac{A_{u^{1}} - \bar{\zeta} A_{\bar{u}^{2}}}{1 + \zeta \bar{\zeta}} \, \dr v^{1} 
    + \frac{A_{u^{2}} + \bar{\zeta} A_{\bar{u}^{1}}}{1 + \zeta \bar{\zeta}} \, \dr v^{2} 
    + \cA_{\zeta} \, \dr \zeta \ , \\
    A^{0,1} & = 
    \frac{A_{\bar{u}^{1}} - \zeta A_{u^{2}}}{1 + \zeta \bar{\zeta}} \, \dr \bar{v}^{1} 
    + \frac{A_{\bar{u}^{2}} + \zeta A_{u^{1}}}{1 + \zeta \bar{\zeta}} \, \dr \bar{v}^{2} 
    + \cA_{\bar{\zeta}} \, \dr \bar{\zeta} \ . 
\end{aligned}
\end{equation}
In these expressions, all of the $\bCP^{1}$-dependence is given explicitly, and $\{ \cA_{\zeta} , \cA_{\bar{\zeta}} \}$ are determined by the constraints $\pd_{\zeta} \vee A = 0$ and $\pd_{\bar{\zeta}} \vee A = 0$. Whilst only the $(0,1)$-form components of $A$ contribute to the sdYM equation on twistor space \eqref{eq:twistorsdYM}, these components are sufficient to capture all of the degrees of freedom in $A$.

On the other hand, let us denote a generic $(0,1)$-form gauge field on twistor space by $\cA$ whose field strength is given by $\cF = \dr \cA + \cA \wedge \cA$. We may impose the constraint $\pd_{\bar{\zeta}} \vee \cA = 0$ by a gauge transformation, provided that $\cA$ is gauge trivial upon restriction to $\bCP^{1}$. We then have
\begin{equation}
    \dr \zeta \wedge \dr v^{1} \wedge \dr v^{2} \wedge \cF = 0 \qquad \Longleftrightarrow \qquad 
    \cF^{0,2} = 0 \ . 
\end{equation}
We will refer to a gauge field which satisfies this equation as \textit{holomorphic}. Given that we are in the gauge $\pd_{\bar{\zeta}} \vee \cA = 0$, the component of this equation given by $\pd_{\bar{\zeta}} \vee \cF^{0,2} = 0$ may be solved explicitly. A generic solution is given by 
\begin{equation}
    \cA = 
    \frac{A_{\bar{u}^{1}} - \zeta A_{u^{2}}}{1 + \zeta \bar{\zeta}} \, \dr \bar{v}^{1} 
    + \frac{A_{\bar{u}^{2}} + \zeta A_{u^{1}}}{1 + \zeta \bar{\zeta}} \, \dr \bar{v}^{2} 
    + \cA_{\bar{\zeta}} \, \dr \bar{\zeta} \ . 
\end{equation}
The components of $A$ appearing in this solution represent generic $\bCP^{1}$-independent functions whose names we have chosen to emphasise the relationship to SDYM. Indeed, following the analysis above, we see that the remaining components of $\cF^{0,2} = 0$ become the sdYM equation for the spacetime gauge field $A$. In summary, the Penrose-Ward transform describes a one-to-one correspondence between solutions to the sdYM equation on $\bR^{4}$, and holomorphic gauge fields on twistor space which are trivial upon restriction to $\bCP^{1}$.

\subsection{Fundamentals of 6dCS theory}

Six-dimensional Chern-Simons (6dCS) theory provides an action for the Penrose-Ward transform. Its equations of motion on twistor space are $\cF^{0,2} = 0$, and it localises to field theories on $\bR^{4}$ whose equations of motion are equivalent to the sdYM equation. In preparation for defining this theory, let us consider the space of $(3,0)$-forms on twistor space. Taking the example from the previous section, its presentation on the northern patch is 
\begin{equation}
    \dr \zeta \wedge \dr v^{1} \wedge \dr v^{2} = - \frac{\dr \tilde{\zeta} \wedge \dr \tilde{v}^{1} \wedge \dr \tilde{v}^{2}}{\tilde{\zeta}^{4}} \ . 
\end{equation}
This $(3,0)$-form appears to be regular on the southern patch, but we see that it has a singularity at $\tilde{\zeta} = 0$. The location and order of these poles have physical significance in the theory, and different setups will lead to different theories on $\bR^{4}$. When defining 6dCS theory, one must specify a choice of $(3,0)$-form on twistor space, which amounts to choosing the location and order of any poles and zeros. The geometry of twistor space dictates that there will always be four more poles than zeros when counted with multiplicity.

In the example above, the $(3,0)$-form is nowhere vanishing and has a single fourth order pole at $\tilde{\zeta} = 0$. This is an interesting model, but let us consider an alternative $(3,0)$-form given by 
\begin{equation}
    \Omega = \frac{\dr \zeta \wedge \dr v^{1} \wedge \dr v^{2}}{\zeta^{2}} = - \frac{\dr \tilde{\zeta} \wedge \dr \tilde{v}^{1} \wedge \dr \tilde{v}^{2}}{\tilde{\zeta}^{2}} \ . 
\end{equation}
This $(3,0)$-form is nowhere vanishing and has two second order poles at $\zeta = 0$ and $\tilde{\zeta} = 0$. Having made a choice of $(3,0)$-form $\Omega$, the 6dCS action is given by 
\begin{equation}
    S_{\text{6dCS}} [\cA] = \frac{1}{2 \pi \rmi} \int_{\bPT} \Omega \wedge \tr \bigg( \cA \wedge \bar{\pd} \cA + \frac{2}{3} \cA \wedge \cA \wedge \cA \bigg) \ . 
\end{equation}
This does not completely define the theory, as we can see by varying the action, 
\begin{equation}
    \delta S_{\text{6dCS}} = \frac{2}{2 \pi \rmi} \int_{\bPT} \Omega \wedge \tr \big( \delta \cA \wedge \cF \big) 
    + \frac{1}{2 \pi \rmi} \int_{\bPT} \bar{\pd} \Omega \wedge \tr \big( \delta \cA \wedge \cA \big) \ . 
\end{equation}
The first term provides the bulk equation of motion $\cF^{0,2} = 0$, and the second term is a boundary contribution at the poles of $\Omega$. This follows from the complex analysis identity 
\begin{equation}
    \pd_{\bar{\zeta}} \bigg( \frac{1}{\zeta} \bigg) = - 2 \pi \rmi \, \delta (\zeta) \ , \qquad 
    \int_{\bCP^{1}} \dr \zeta \wedge \dr \bar{\zeta} \, \delta(\zeta) \, f(\zeta) = f(0) \ . 
\end{equation}
Since our choice of $(3,0)$-form contains second order poles, the boundary term includes derivatives of delta-functions, and higher order poles would produce higher order derivatives. The contribution from the southern patch may be explicitly evaluated to give 
\begin{equation}
\begin{aligned}
    \frac{1}{2 \pi \rmi} \int_{\bPT} \bar{\pd} \Omega \wedge \tr \big( \delta \cA \wedge \cA \big) 
    = \int_{\bPT} \dr \zeta \wedge \dr \bar{\zeta} \, \delta(\zeta) \, \pd_{\zeta} \bigg[ \dr v^{1} \wedge \dr v^{2} \wedge \tr \big( \delta \cA \wedge \cA \big) \bigg] \qquad & \\ 
    = \int_{\bR^{4}} \bigg[ \, \mu \wedge \tr \big( \delta \cA \wedge \cA \big) \big\vert_{\zeta = 0} + \dr u^{1} \wedge \dr u^{2} \wedge \pd_{\zeta} \tr \big( \delta \cA \wedge \cA \big) \big\vert_{\zeta = 0} \bigg] & \ . 
\end{aligned}
\end{equation}
As in the previous section, the 2-form $\mu = \dr u^{1} \wedge \dr \bar{u}^{1} + \dr u^{2} \wedge \dr \bar{u}^{2}$ is proportional to the K{\"a}hler form on $\bR^{4}$. Boundary conditions must be imposed on the gauge field such that the boundary terms in the variation vanish. Notably, these conditions may generically constrain not only the value of the gauge field at the poles, but also its $\bCP^{1}$-derivatives. In this case, it is sufficient to impose the boundary conditions 
\begin{equation}
    \cA \big\vert_{\zeta = 0} = 0 \ , \qquad 
    \cA \big\vert_{\tilde{\zeta} = 0} = 0 \ . 
\end{equation}
The second of these conditions follows from a similar analysis on the northern patch.

The Chern-Simons 3-form is invariant under infinitesimal gauge transformations up to boundary terms, which must be considered carefully in the present context. Infinitesimal gauge transformations act on the gauge field as 
\begin{equation}
    \delta \cA = \bar{\pd} \epsilon + [ \cA , \epsilon ] \ . 
\end{equation}
Generic transformations do not preserve the action whose variation is given by 
\begin{equation}
    \delta S_{\text{6dCS}} = \frac{1}{2 \pi \rmi} \int_{\bPT} \bar{\pd} \Omega \wedge \tr \big( \cA \wedge \bar{\pd} \epsilon \big) \ . 
\end{equation}
This boundary term may be computed explicitly, as before, though some simplifications occur due to the boundary conditions on $\cA$. The contribution from the southern patch is 
\begin{equation}
    \frac{1}{2 \pi \rmi} \int_{\bPT} \bar{\pd} \Omega \wedge \tr \big( \cA \wedge \bar{\pd} \epsilon \big) = \int_{\bR^{4}} \dr u^{1} \wedge \dr u^{2} \wedge \tr \big( \pd_{\zeta} \cA \wedge \bar{\pd} \epsilon \big) \big\vert_{\zeta = 0} \ . 
\end{equation}
This term, along with its counterpart on the northern patch, will vanish if the gauge transformation parameter satisfies 
\begin{equation} \label{eq:gaugeBC}
    \bar{\pd} \epsilon \big\vert_{\zeta = 0} = 0 \ , \qquad 
    \bar{\pd} \epsilon \big\vert_{\tilde{\zeta} = 0} = 0 \ . 
\end{equation}
At this point, it is important to linger on the physical status of these symmetries. In general, gauge symmetries describe redundancies of a theory that should be removed by imposing a gauge fixing condition. Crucially, physical observables must not depend on this choice of gauge fixing. On the other hand, a given theory may also admit physical symmetries which are accompanied by a conserved charge according to Noether's theorem. States in such a theory organise themselves into representations of the physical symmetries, distinguished by differing values of the conserved charge. One way to determine whether a symmetry is physical or gauge is via its conserved charge: the charge associated to a gauge symmetry always vanishes.

With this in mind, let us consider the infinitesimal transformations satisfying the boundary conditions \eqref{eq:gaugeBC}. Those transformations which act trivially at the poles, but non-trivially in the bulk, are the genuine local gauge transformations of Chern-Simons theory. To understand the transformations which act non-trivially at the poles, it is important to accurately identify the boundary degrees of freedom in this theory. As we have seen, the boundary terms typically depend on both the value of the gauge field and its $\bCP^{1}$-derivative due to the second order poles in $\Omega$. Similarly, the infinitesimal transformations can act non-trivially at the poles in two different ways. Either the value of $\epsilon$ is non-vanishing at the poles, or it has a non-vanishing $\bCP^{1}$-derivative. It turns out that the latter are gauge symmetries, whilst the former are physical symmetries with associated conservation laws given by 
\begin{equation} \label{eq:6dCSconservation}
    \dr u^{1} \wedge \dr u^{2} \wedge \bar{\pd} (\pd_{\zeta} \cA) \big\vert_{\zeta = 0} = 0 \ , \qquad 
    \dr \bar{u}^{1} \wedge \dr \bar{u}^{2} \wedge \bar{\pd} (\pd_{\tilde{\zeta}} \cA) \big\vert_{\tilde{\zeta} = 0} = 0 \ . 
\end{equation}
In these expressions, the Noether currents are $\pd_{\zeta} \cA \big\vert_{\zeta = 0}$ and $\pd_{\tilde{\zeta}} \cA \big\vert_{\tilde{\zeta} = 0}$ which are generically non-vanishing. These currents are not conserved in the usual sense, but rather satisfy a holomorphicity condition. This follows from the fact that general solutions to the boundary conditions \eqref{eq:gaugeBC} are not global transformations, but semi-local transformations with partial dependence on $\bR^{4}$.

The exponentiation of these infinitesimal transformations are the finite gauge transformations which act on the gauge field as 
\begin{equation}
    \cA \mapsto \cA^{g} = g^{-1} \cA g + g^{-1} \bar{\pd} g \ . 
\end{equation}
These preservere the space of solutions to the bulk equations of motion $\cF^{0,2} = 0$ as the field strength transforms as $\cF \mapsto g^{-1} \cF g$. To understand the need for boundary conditions, we should consider the transformation of the action, 
\begin{equation}
    S_{\text{6dCS}} [\cA] \mapsto S_{\text{6dCS}} [\cA] + \frac{1}{2 \pi \rmi} \int_{\bPT} \bar{\pd} \Omega \wedge \tr \big( \cA \wedge \bar{\pd} g g^{-1} \big) 
    - \frac{1}{6 \pi \rmi} \int_{\bPT} \Omega \wedge \tr \big( g^{-1} \bar{\pd} g \big)^{3} \ . 
\end{equation}
Using the boundary conditions on $\cA$, one may see that the second term vanishes if we impose boundary conditions of the gauge transformations given by 
\begin{equation}
    g^{-1} \bar{\pd} g \big\vert_{\zeta = 0} = 0 \ , \qquad 
    g^{-1} \bar{\pd} g \big\vert_{\tilde{\zeta} = 0} = 0 \ . 
\end{equation}
These are very similar to the boundary conditions we imposed on the infinitesimal gauge transformations \eqref{eq:gaugeBC}. The third term is most easily understood by introducing an extension of $g$ over the 7-manifold $\bPT \times [0,1]$ which agrees with $g$ on $\bPT \times \{ 1 \}$ and with the trivial map on $\bPT \times \{ 0 \}$. Denoting this extension by $\tilde{g}$, we define the Wess-Zumino (WZ) 2-form by 
\begin{equation}
    \text{WZ} [g] = \frac{1}{3} \int_{[0,1]} \tr \big( \tilde{g}^{-1} \dr \tilde{g} \wedge \tilde{g}^{-1} \dr \tilde{g} \wedge \tilde{g}^{-1} \dr \tilde{g} \big) \ . 
\end{equation}
To relate this to the third term in the gauge transformation of the action, we first note that we may replace the Dobeault operators $\bar{\pd}$ by standard exterior derivatives whenever they are wedged against $\Omega$. Then, we leverage Stokes's theorem and the extension described above to write the integral over the 6-manifold $\bPT$ as a surface integral over the 7-manifold $\bPT \times [0,1]$. Since the $\tilde{g}$-dependent 3-form is closed, the overall exterior derivative can only act on $\Omega$, giving us the identity 
\begin{equation}
    \frac{1}{3} \int_{\bPT} \Omega \wedge \tr \big( g^{-1} \bar{\pd} g \wedge g^{-1} \bar{\pd} g \wedge g^{-1} \bar{\pd} g \big) = \int_{\bPT} \bar{\pd} \Omega \wedge \text{WZ} [g] \ . 
\end{equation}
This means that we can write the gauge transformation of the action succinctly as 
\begin{equation}
    S_{\text{6dCS}} [\cA^{g}] = S_{\text{6dCS}} [\cA] + \frac{1}{2 \pi \rmi} \int_{\bPT} \bar{\pd} \Omega \wedge \bigg[ \tr \big( \cA \wedge \bar{\pd} g g^{-1} \big) 
    - \text{WZ} [g] \bigg] \ . 
\end{equation}
Explicitly computing the localisation of the WZ-term, one finds that the boundary conditions we have already imposed on the finite transformations are sufficient to make it vanish. As with the infinitesimal transformations, the finite transformations include both a gauge and physical component, distinguished by their Noether charges.

In addition to these transformations of the fields, we should also consider diffeomorphisms of twistor space. The action is not invariant under general diffeomorphisms, only under those which preserve $\Omega$. This includes all translations along $\bR^{4}$, but does not include all SO$(4)$ rotations of spacetime. There is an SU$(2)$ subgroup of rotations preserving $\Omega$ which is generated by 
\begin{equation}
\begin{aligned}
    R_{1}^{\prime \prime} = & \frac{\rmi}{2} \big( u^{1} \pd_{u^{1}} - \bar{u}^{1} \pd_{\bar{u}^{1}} - u^{2} \pd_{u^{2}} + \bar{u}^{2} \pd_{\bar{u}^{2}} \big) \ , \\ 
    R_{2}^{\prime \prime} = & \frac{\rmi}{2} \big( u^{2} \pd_{u^{1}} - \bar{u}^{2} \pd_{\bar{u}^{1}} + u^{1} \pd_{u^{2}} - \bar{u}^{1} \pd_{\bar{u}^{2}} \big) \ , \\ 
    R_{3}^{\prime \prime} = & \frac{1}{2} \big( u^{2} \pd_{u^{1}} + \bar{u}^{2} \pd_{\bar{u}^{1}} - u^{1} \pd_{u^{2}} - \bar{u}^{1} \pd_{\bar{u}^{2}} \big) \ . 
\end{aligned}
\end{equation}
These act trivially on the $\bCP^{1}$-coordinate $\zeta$, but non-trivially on the holomorphic coordinates $\{ v^{1} , v^{2} \}$ of twistor space. There is also a U$(1)$ diffeomorphism acting on $\bR^{4}$ and $\bCP^{1}$ generated by 
\begin{equation}
    R_{0}^{\prime \prime} = \frac{\rmi}{2} \big( u^{1} \pd_{u^{1}} - \bar{u}^{1} \pd_{\bar{u}^{1}} + u^{2} \pd_{u^{2}} - \bar{u}^{2} \pd_{\bar{u}^{2}} \big) + \rmi \zeta \pd_{\zeta} \ . 
\end{equation}
This commutes with the SU$(2)$ subgroup of rotations, and together they form the $\text{U}(2) \subset \text{SO}(4)$ which preserves the K{\"a}hler form on $\bR^{4}$.

\subsection{Localisation to spacetime}
\begin{equation*}
\begin{tikzpicture}
\node at (-2,1) {\textbf{6dCS}};
\node at (-2,-1) {\textbf{4dCS}};
\node at (2,1) {\textbf{4dIFT}};
\node at (2,-1) {\textbf{2dIFT}};
\draw[->, draw=tab10red, very thick] (-1,1) -- (1,1);
\draw[->, very thick] (-1,-1) -- (1,-1);
\draw[->, very thick, decorate, decoration={snake, segment length=12.5pt, amplitude=2pt}] (-2,0.75) -- (-2,-0.75);
\draw[->, very thick, decorate, decoration={snake, segment length=12.5pt, amplitude=2pt}] (2,0.75) -- (2,-0.75);
\end{tikzpicture}
\end{equation*}

Next, we would like to show that 6dCS theory localises to a field theory on spacetime whose equations of motion are equivalent to the sdYM equation. In particular, unlike a Kaluza-Klein reduction in which infinitely many modes are discarded from the theory, a finite number of fields on $\bR^{4}$ capture all of the physical degrees of freedom in 6dCS theory. This is analogous to the localisation of 4dCS theory which is presented in section \ref{sec:4dCS}.
In that context, we saw that the bulk degrees of freedom in the theory were gauge trivial, and the only physical degrees of freedom arose from boundary conditions imposed at poles in $\bCP^{1}$. 
The same argument applies to 6dCS theory, where the role of the boundary is played by the poles in $\Omega$. We will refer to the degrees of freedom living at the poles as \textit{edge modes}, and they will become the fundamental fields of the 4d theory.

Now that we have the idea in mind, let us explicitly localise 6dCS theory to derive the theory on spacetime. It is helpful to introduce a field redefinition which separates the bulk gauge field from the edge modes. To this end, we introduce two new fundamental fields $\cA^{\prime}$ and $\hat{g}$ which are related to $\cA$ by 
\begin{equation}
    \cA = \cA^{\prime \hat{g}} = \hat{g}^{-1} \cA^{\prime} \hat{g} + \hat{g}^{-1} \bar{\pd} \hat{g} \ . 
\end{equation}
This new parameterisation is partially redundant and introduces an \textit{internal} gauge symmetry which acts as 
\begin{equation}
    \cA^{\prime} \mapsto \cA^{\prime \check{h}} \ , \qquad 
    \hat{g} \mapsto \check{h}^{-1} \hat{g} \ . 
\end{equation}
Notice that the original field $\cA$ is invariant under this transformation meaning that it trivially preserves the action. Assuming that $\cA^{\prime}$ is trivial upon restriction to $\bCP^{1}$, we can leverage this internal symmetry to impose the gauge fixing constraint 
\begin{equation}
    \pd_{\bar{\zeta}} \vee \cA^{\prime} = 0 \ . 
\end{equation}
This constraint is familiar from our discussion of the Penrose-Ward transform in appendix \ref{sec:PWtransform}, and it will always be applied when localising 6dCS theory. By comparison, we will shortly impose some additional gauge fixing conditions on the edge mode $\hat{g}$, but these will vary from model to model depending on the choice of $\Omega$ and boundary conditions.

Before imposing these additional gauge fixing conditions, let us revisit the action of 6dCS theory. In the new variables, it is given by 
\begin{equation}
    S_{\text{6dCS}} [\cA^{\prime} , \hat{g}] = S_{\text{6dCS}} [\cA^{\prime}] + \frac{1}{2 \pi \rmi} \int_{\bPT} \bar{\pd} \Omega \wedge \bigg[ \tr \big( \cA^{\prime} \wedge \bar{\pd} \hat{g} \hat{g}^{-1} \big) 
    - \text{WZ} [\hat{g}] \bigg] \ . 
\end{equation}
The field $\hat{g}$ only appears in the action against $\bar{\pd} \Omega$, justifying its title of edge mode. In particular, while one might initially think of $\hat{g}$ as defined over the whole manifold $\bPT$, we see that it only enters into the action through the fields 
\begin{equation}
    \hat{g} \big\vert_{\zeta = 0} = g \ , \qquad 
    \hat{g}^{-1} \pd_{\zeta} \hat{g} \big\vert_{\zeta = 0} = \phi \ , \qquad 
    \hat{g} \big\vert_{\tilde{\zeta} = 0} = \tilde{g} \ , \qquad 
    \hat{g}^{-1} \pd_{\tilde{\zeta}} \hat{g} \big\vert_{\tilde{\zeta} = 0} = \tilde{\phi} \ . 
\end{equation}
These 4d fields capture all of the degrees of freedom in the edge mode $\hat{g}$, where the $\bCP^{1}$-derivatives appear because of the second order poles in $\Omega$.

However, some of these degrees of freedom are non-physical due to the presence of gauge symmetries. For example, compatible with the gauge fixing condition $\pd_{\bar{\zeta}} \vee \cA^{\prime} = 0$, we may apply internal gauge transformations satisfying $\pd_{\bar{\zeta}} \check{h} = 0$. This provides sufficient freedom to gauge fix the value of $\hat{g}$ at one point on $\bCP^{1}$ to the identity, for example $\tilde{g} = \id$. In addition, we also have access to the original gauge transformations which we will refer to as \textit{external} symmetries. These act on the new variables as 
\begin{equation}
    \cA^{\prime} \mapsto \cA^{\prime} \ , \qquad 
    \hat{g} \mapsto \hat{g} \hat{h} \ . 
\end{equation}
Notice that our field redefinition has conveniently decoupled the unconstrained bulk gauge transformations from the residual symmetries acting on the boundary degrees of freedom. As we emphasised in the previous section, the parameter $\hat{h}$ contains both gauge symmetries and semi-local physical symmetries. The gauge symmetries must be trivial at the poles, but may have non-trivial $\bCP^{1}$-derivatives. These allow us to impose the constraints $\phi = 0$ and $\tilde{\phi} = 0$. In summary, the gauge fixing conditions imposed on the edge mode are 
\begin{equation}
    \hat{g} \big\vert_{\zeta = 0} = g \ , \qquad 
    \hat{g}^{-1} \pd_{\zeta} \hat{g} \big\vert_{\zeta = 0} = 0 \ , \qquad 
    \hat{g} \big\vert_{\tilde{\zeta} = 0} = \id \ , \qquad 
    \hat{g}^{-1} \pd_{\tilde{\zeta}} \hat{g} \big\vert_{\tilde{\zeta} = 0} = 0 \ . 
\end{equation}
This exhausts the gauge symmetries of the theory, and the semi-local symmetries act on the surviving degrees of freedom as 
\begin{equation}
    \cA^{\prime} \mapsto \cA^{\prime h_{\ell}} \ , \qquad 
    g \mapsto h_{\ell}^{-1} g h_{r} \ , \qquad 
    \pd h_{\ell} = 0 \ , \qquad 
    \bar{\pd} h_{r} = 0 \ . 
\end{equation}
In this expressions, the transformation parameters $h_{\ell}$ and $h_{r}$ only depend on $\bR^{4}$, and we have written the semi-local constraints in terms of the complex coordinates $\{ u^{1} , \bar{u}^{1} , u^{2} , \bar{u}^{2} \}$. In general, when we write Dobeault operators acting on fields on $\bR^{4}$, they are defined with respect to the complex structure in which $\{ u^{1}, u^{2} \}$ are holomorphic coordinates. This coincides with the complex structure on twistor space at $\zeta = 0$, and defines the opposite complex structure (with holomorphic and anti-holomorphic exchanged) when compared to $\tilde{\zeta} = 0$. For example, we can write the boundary conditions on the infinitesimal transformations as 
\begin{equation}
    \bar{\pd} \epsilon \big\vert_{\zeta = 0} = \bar{\pd} \big( \epsilon \big\vert_{\zeta = 0} \big) = 0 \ , \qquad 
    \bar{\pd} \epsilon \big\vert_{\tilde{\zeta} = 0} = \pd \big( \epsilon \big\vert_{\tilde{\zeta} = 0} \big) = 0 \ . 
\end{equation}
We expect these residual symmetries to descend to semi-local symmetries of the theory on spacetime.

The next step in the localisation procedure is to solve the $\delta \cA^{\prime}$ bulk equations of motion. Taking into account the constraint $\pd_{\bar{\zeta}} \vee \cA^{\prime} = 0$, this equation of motion is given by 
\begin{equation}
    \Omega \wedge \cL_{\bar{\zeta}} \cA^{\prime} = 0 \ . 
\end{equation}
We have already seen the general solution in our review of the Penrose-Ward transform \S \ref{sec:PWtransform}, but here we will present a more detailed derivation. When writing $\cA^{\prime}$ in components, it is natural to work in the basis of $(0,1)$-forms given by $\{ \dr \bar{\zeta} , \dr \bar{v}^{1} , \dr \bar{v}^{2} \}$. Unfortunately, this basis has some drawbacks and is not particularly well adapted for localising to spacetime. Firstly, the $(0,1)$-forms along $\bR^{4}$ also have legs along $\bCP^{1}$, meaning that the constraint $\pd_{\bar{\zeta}} \vee \cA^{\prime} = 0$ relates the various components rather than simply setting one to zero. Secondly, the $(0,1)$-forms along $\bR^{4}$ are not invariant along $\bCP^{1}$, meaning that the Lie derivative acts on both the components and the basis forms. Of course, one may overcome these obstacles and complete the calculation in this basis, but here we will present an alternative approach.

The localisation is made easier by working in a basis of forms which does not suffer from these downsides. It turns out that asking for invariance along $\bCP^{1}$ is too restrictive, but we can find some $(0,1)$-forms $\{ \bar{\theta}^{1} , \bar{\theta}^{2} \}$ satisfying 
\begin{equation}
    \pd_{\bar{\zeta}} \vee \bar{\theta}^{1} = 0 \ , \qquad 
    \Omega \wedge \cL_{\bar{\zeta}} \bar{\theta}^{1} = 0 \ , \qquad 
    \pd_{\bar{\zeta}} \vee \bar{\theta}^{2} = 0 \ , \qquad 
    \Omega \wedge \cL_{\bar{\zeta}} \bar{\theta}^{2} = 0 \ . 
\end{equation}
These $(0,1)$-forms are explicitly given by 
\begin{equation}
    \bar{\theta}^{1} = \frac{\dr \bar{u}^{1} - \bar{\zeta} \dr u^{2}}{1 + \zeta \bar{\zeta}} \ , \qquad
    \bar{\theta}^{2} = \frac{\dr \bar{u}^{2} + \bar{\zeta} \dr u^{1}}{1 + \zeta \bar{\zeta}} \ . 
\end{equation}
Working in the basis of $(0,1)$-forms given by $\{ \dr \bar{\zeta} , \bar{\theta}^{1} , \bar{\theta}^{2} \}$, the constraint $\pd_{\bar{\zeta}} \vee \cA^{\prime} = 0$ simply implies that the coefficient of $\dr \bar{\zeta}$ vanishes. Furthermore, the bulk equation of motion implies that the coefficients of $\bar{\theta}^{1}$ and $\bar{\theta}^{2}$ are holomorphic functions of $\bCP^{1}$. Expanding these holomorphic coefficients as polynomials in $\zeta$, the requirement that $\cA^{\prime}$ is finite everywhere on $\bCP^{1}$ tells us that they may be at most degree 1. The general solution to the bulk equation of motion is therefore given by 
\begin{equation}
    \cA^{\prime} = 
    (A_{\bar{u}^{1}} - \zeta A_{u^{2}}) \, \bar{\theta}^{1} 
    + (A_{\bar{u}^{2}} + \zeta A_{u^{1}}) \, \bar{\theta}^{2} \ . 
\end{equation}
This completely specifies the $\bCP^{1}$-dependence of $\cA^{\prime}$, and agrees with the result presented in appendix \ref{sec:PWtransform}.

Having found the $\bCP^{1}$-dependence of $\cA^{\prime}$, we could now compute the integral over $\bCP^{1}$ in the action arriving at a theory on $\bR^{4}$. As it stands, that theory would depend on the surviving edge mode $g$ and the 4d gauge field $A$. However, we have yet to take into account the boundary conditions of 6dCs theory. These allow us to solve for $A$ in terms of $g$, landing on a theory with a single fundamental field. We expect the equations of motion of this theory to be equivalent to the sdYM equation for $A$.

The boundary conditions on $\cA$ may be converted into conditions on $\cA^{\prime}$ and $\hat{g}$ as 
\begin{equation}
    \cA^{\prime} \big\vert_{\zeta = 0} = \bar{\pd} \hat{g} \hat{g}^{-1} \big\vert_{\zeta = 0} \ , \qquad 
    \cA^{\prime} \big\vert_{\tilde{\zeta} = 0} = \bar{\pd} \hat{g} \hat{g}^{-1} \big\vert_{\tilde{\zeta} = 0} \ . 
\end{equation}
Solving these for the components of $\cA^{\prime}$ gives 
\begin{equation}
    A_{u^{1}} = 0 \ , \qquad 
    A_{\bar{u}^{1}} = \pd_{\bar{u}^{1}} g g^{-1} \ , \qquad 
    A_{u^{2}} = 0 \ , \qquad 
    A_{\bar{u}^{2}} = \pd_{\bar{u}^{2}} g g^{-1} \ .
\end{equation}
Let us consider the corresponding 4d gauge field $A$. In the complex structure on $\bR^{4}$, this form of the gauge field may be summarised as $A^{1,0} = 0$ and $A^{0,1} = - \bar{\pd} g g^{-1}$. This connection identically solves two of the sdYM equations, 
\begin{equation}
    F^{2,0} = 0 \ , \qquad 
    F^{0,2} = 0 \ . 
\end{equation}
The first of these is solved trivially, and the second is solved due to the Maurer-Cartan identity. We expect the final sdYM equation to coincide with the equation of motion of the 4d theory, 
\begin{equation}
    \mu \wedge F = 0 \quad \Longleftrightarrow \quad 
    \mu \wedge \pd ( \bar{\pd} g g^{-1} ) = 0 \ . 
\end{equation}
This parameterisation of a self-dual connection is well-known, and the field $g$ is often referred to as \textit{Yang's matrix}. In this context, final sdYM equation is known as Yang's equation.

To find the action of the 4d theory, we should substitute our solution for $\cA^{\prime}$ into the 6dCS action and compute the integral over $\bCP^{1}$. Recall that the 6dCS action is given by 
\begin{equation}
    S_{\text{6dCS}} [\cA^{\prime} , \hat{g}] = S_{\text{6dCS}} [\cA^{\prime}] + \frac{1}{2 \pi \rmi} \int_{\bPT} \bar{\pd} \Omega \wedge \bigg[ \tr \big( \cA^{\prime} \wedge \bar{\pd} \hat{g} \hat{g}^{-1} \big) 
    - \text{WZ} [\hat{g}] \bigg] \ . 
\end{equation}
Since the solution to the bulk equation of motion obeys $\pd_{\bar{\zeta}} \vee \cA^{\prime} = 0$ and $\Omega \wedge \cL_{\bar{\zeta}} \cA^{\prime} = 0$, the integrand in the first term cannot saturate the $\dr \bar{\zeta}$ leg and therefore vanishes. It remains to compute the integral over $\bCP^{1}$ in the second term, which we will do piece by piece. Since we have gauge fixed all of the degrees of freedom in the edge mode at the north pole, this integral will only receive contributions from the south pole. The first of these is given by 
\begin{equation}
    \int_{\bR^{4}} \bigg[ \, \mu \wedge \tr \big( \cA^{\prime} \wedge \bar{\pd} \hat{g} \hat{g}^{-1} \big) \big\vert_{\zeta = 0} + \dr u^{1} \wedge \dr u^{2} \wedge \pd_{\zeta} \tr \big( \cA^{\prime} \wedge \bar{\pd} \hat{g} \hat{g}^{-1} \big) \big\vert_{\zeta = 0} \bigg] \ . 
\end{equation}
This expression may be simplified with a couple of observations. Firstly, the value of $\cA^{\prime}$ at the south pole is $\cA^{\prime} \vert_{\zeta = 0} = A^{0,1} = - \bar{\pd} g g^{-1}$. Similarly, since the complex structure on twistor space agrees with that on spacetime at $\zeta = 0$, we have $\bar{\pd} \hat{g} \hat{g}^{-1} \vert_{\zeta = 0} = \bar{\pd} g g^{-1}$. Inspecting the first term, we see that the integrand would be a $(1,3)$-form on $\bR^{4}$. This does not exist, so this term must vanish.

Secondly, the $\bCP^{1}$-derivative of $\cA^{\prime}$ vanishes at the south pole $\pd_{\zeta} \cA^{\prime} \vert_{\zeta = 0} = 0$. The $\bCP^{1}$-derivative of the edge mode also vanishes due to our choice of gauge fixing, so this derivative may only act on the $\bCP^{1}$-dependence inside the twistor space Dobeault operator $\bar{\pd}$. In the basis of $(0,1)$-forms $\{ \dr \bar{\zeta} , \bar{\theta}^{1} , \bar{\theta}^{2} \}$, this operator acts on functions as 
\begin{equation}
    \bar{\pd} f = \pd_{\bar{\zeta}} f \, \dr \bar{\zeta} 
    + (\pd_{\bar{u}^{1}} f - \zeta \, \pd_{u^{2}} f) \, \bar{\theta}^{1} 
    + (\pd_{\bar{u}^{2}} f + \zeta \, \pd_{u^{1}} f) \, \bar{\theta}^{2} \ . 
\end{equation}
This allows us to compute the contribution of the second term which is 
\begin{equation}
    \int_{\bR^{4}} \dr u^{1} \wedge \dr u^{2} \wedge \pd_{\zeta} \tr \big( \cA^{\prime} \wedge \bar{\pd} \hat{g} \hat{g}^{-1} \big) \big\vert_{\zeta = 0} 
    = \frac{1}{2} \int_{\bR^{4}} \tr \big( g^{-1} \dr g \wedge {\star} g^{-1} \dr g \big) \ . 
\end{equation}
This is the standard kinetic term of the principal chiral model (PCM). Using some identities from K{\"ahler} geometry, this may also be written as 
\begin{equation}
    \frac{1}{2} \int_{\bR^{4}} \tr \big( g^{-1} \dr g \wedge {\star} g^{-1} \dr g \big) 
    = \int_{\bR^{4}} \mu \wedge \tr \big( g^{-1} \pd g \wedge g^{-1} \bar{\pd} g \big) \ . 
\end{equation}

Turning to the WZ-term, this computation is very direct. Since the $\bCP^{1}$-derivative of the edge mode has been fixed, we only pick up a term of the form $\omega \wedge \text{WZ}[g]$. Bringing these pieces together, the action of the 4d theory is 
\begin{equation}
    S_{\text{4dWZW}} [g] = \frac{1}{2} \int_{\bR^{4}} \tr \big( g^{-1} \dr g \wedge {\star} g^{-1} \dr g \big) 
    - \int_{\bR^{4}} \mu \wedge \text{WZ}[g] \ . 
\end{equation}
This theory is known as the four-dimensional Wess-Zumino-Witten (4dWZW) model. As expected, the equations of motion coincide with the final sdYM equation, 
\begin{equation}
    \delta S_{\text{4dWZW}} = 0 \quad \Longleftrightarrow \quad 
    \mu \wedge \pd (\bar{\pd} g g^{-1}) = 0 \ . 
\end{equation}
In addition, the action is invariant under two semi-local symmetries acting as 
\begin{equation}
    g \mapsto h_{\ell}^{-1} g h_{r} \ , \qquad 
    \pd h_{\ell} = 0 \ , \qquad 
    \bar{\pd} h_{r} = 0 \ . 
\end{equation}
These are directly inherited from the residual symmetries of 6dCS theory which satisfy the boundary conditions. The associated conservation laws take the form of holomorphicity conditions, 
\begin{equation}
    \mu \wedge \pd (\bar{\pd} g g^{-1}) = 0 \ , \qquad 
    \mu \wedge \bar{\pd} (g^{-1} \pd g) = 0 \ . 
\end{equation}
One may show that these agree with the conservation laws derived from 6dCS theory \eqref{eq:6dCSconservation} when substituting in the field redefinition and solution to the bulk equation of motion.

Considering the spacetime symmetries, we see that only those diffeomorphisms which preserve the K{\"a}hler form will leave the action invariant. This includes all translations of $\bR^{4}$ and a subgroup $\text{U}(2) \subset \text{SO}(4)$ of the rotations. This matches the 6dCS analysis.

\section{Symmetric space PCM from 6dCS theory}
\label{app:symm_PCM}

In the main body of this paper, we recovered the 2dIFT related to stationary axisymmetric gravity, which takes the form of a symmetric-space $\sigma$-model with a spacetime-dependent coupling. This involved imposing a translational, a rotational, and a discrete reduction, either to go from the 4dWZW model to the 2dIFT, or from 6dCS to 4dCS theory. 
The rotational reduction is the origin of the spacetime-dependent coupling, while the discrete reduction brings the target space to the symmetric-space $G / G_{0}$ rather than $G$. 

In this appendix, we instead consider the case where both reduction vectors are translational, isolating the role of the discrete reduction in our story. 
Starting from the 4dWZW model, we recover the 2d principal chiral model (PCM) without a WZ term. This reduction was first performed in \cite{Bittleston:2020hfv, Penna:2020uky} but we introduce the additional ingredient of a discrete reduction involving a $\bZ_{2}$-automorphism of the Lie algebra. 
The addition of this discrete reduction will reduce the target space of the PCM from the full group $G$ to the symmetric space $G / G_{0}$. Here, $G_{0} \subset G$ is the subgroup which is preserved by the $\bZ_{2}$-automorphism. 
This reduction may then be lifted to twistor space and applied to 6dCS theory. We will show that this reproduces the branch cut and equivariance condition discussed in \cite[\S 11]{Costello:2019tri} and known to recover the symmetric-space PCM.

Let us start with the 4dWZW model whose action is 
\begin{equation}
    S_{\text{4dWZW}} = \frac{1}{2} \int_{\bR^{4}} \tr \big( g^{-1} \dr g \wedge {\star} g^{-1} \dr g \big) 
    - \int_{\bR^{4}} \mu \wedge \text{WZ}[g] \ . 
\end{equation}
The second term contains a 2-form $\mu = \dr u^{1} \wedge \dr \bar{u}^{1} + \dr u^{2} \wedge \dr \bar{u}^{2}$ which is proportional to the K{\"a}hler form on $\bR^{4}$ equipped with the Euclidean metric. 
We would like to apply a two-dimensional reduction by vector fields $X$ and $Y$, landing on the 2dPCM whose action is 
\begin{equation}
    S_{\text{2dPCM}} = \frac{1}{2} \int_{\bR^{2}} \tr \big( g^{-1} \dr g \wedge {\star} g^{-1} \dr g \big) \ . 
\end{equation}
In order for the 2d theory to contain no WZ-term, we must choose the vector fields such that $(X \wedge Y) \vee \mu = 0$. Were one to compactify the directions corresponding to these vector fields, this condition would state that the resulting 2-torus has zero K{\"a}hler volume \cite{Costello:2021}. We can satisfy this condition with two translational vector fields 
\begin{equation}
    X = \pd_{u^{1}} - \pd_{\bar{u}^{2}} \ , \qquad 
    Y = \pd_{u^{2}} - \pd_{\bar{u}^{1}} \ . 
\end{equation}
Coordinates on the quotient space must be invariant under these vector fields, and these may be provided by $\xi = u^{1} + \bar{u}^{2}$ and $\bar{\xi} = \bar{u}^{1} + u^{2}$. Applying this reduction to the 4dWZW model results in the 2dPCM as desired.

Next, we would like to supplement this reduction with a discrete reduction. So that we do not modify the geometry of the quotient space, we will choose a discrete action which leaves the quotient space invariant. In addition, this discrete reduction must be a symmetry of the 4dWZW model, naively meaning it must preserve the 2-form $\mu$. In fact, it is enough for this symmetry to preserve the 2-form $\mu$ up to a sign, provided that we also transform the fundamental field as $g \mapsto g^{-1}$. Such a discrete transformation is provided by 
\begin{equation}
    \sigma : (u^{1}, u^{2}) \mapsto (\bar{u}^{2}, \bar{u}^{1}) \ . 
\end{equation}
In order to implement the transformation of the fundamental field, we will introduce a $\bZ_{2}$-automorphism of the Lie algebra, 
\begin{equation}
    \eta : \fg \to \fg \ , \qquad \eta^{2} = \id \ . 
\end{equation}
The total reduction will then be a combination of the discrete spacetime action and this Lie algebra automorphism exponentiated to a group action $g \mapsto \eta(g)$. 
To agree with the transformation specified above, the fundamental field must satisfy $\eta(g) = g^{-1}$, which we thus impose as a constraint. This restricts the target space of the 2dPCM from the full group $G$ to the symmetric space $G / G_{0}$ where $G_{0} \subset G$ is the subgroup preserved by the $\bZ_{2}$-automorphism.

Having outlined the reduction of the 4dWZW model, we may now lift this to a reduction of twistor space and then apply it to 6dCS theory. Firstly, let us study the geometry of the reduced twistor correspondence. The vector fields lift trivially to $\bPS$ meaning that the reduced correspondence space $M_{4} = \bCP^{1} \times \bR^{2}$ may be provided coordinates by $\{ \zeta , \xi , \bar{\xi} \}$. The projection of these vector fields to twistor space $\bPT$ is given by 
\begin{equation}
    X^{\prime} = (1 + \zeta) \pd_{v^{1}} \ , \qquad 
    Y^{\prime} = (1 - \zeta) \pd_{v^{2}} \ . 
\end{equation}
We therefore identify the reduced twistor space as $\bCP^{1}$ with the invariant spectral parameter $\zeta$.

Turning to the discrete reduction, this lifts to the correspondence space as 
\begin{equation}
    \sigma^{\prime \prime} : (u^{1}, u^{2}, \zeta) \mapsto (\bar{u}^{2}, \bar{u}^{1}, \zeta^{-1}) \ . 
\end{equation}
Furthermore, the action of this discrete transformation on twistor space is 
\begin{equation}
    \sigma^{\prime} : (\zeta , v^{1} , v^{2}) \mapsto \bigg( \frac{1}{\zeta} , -\frac{v^{1}}{\zeta} , \frac{v^{2}}{\zeta} \bigg) \ . 
\end{equation}
This is a holomorphic map which preserves the $(3,0)$-form $\Omega$ associated to the 4dWZW model. The key feature of this discrete transformation is its action on the spectral parameter $\zeta$. To understand the implications of this discrete reduction in the 4dCS theory, it is helpful to introduce another spectral parameter which is invariant under the discrete transformation, 
\begin{equation}
    W = \frac{1}{2} \big( \zeta + \zeta^{-1} \big) \ . 
\end{equation}
This is known as the Joukowsky transform and the $\zeta$-plane is a double covering of the $W$-plane. We can see this by considering the inverse relationship 
\begin{equation}
    \zeta = W + \sqrt{W^{2} - 1} \ . 
\end{equation}
For each value of $W$, there are two values of $\zeta$, except for at the points $W = \pm 1$. These special points are the ends of a branch cut in the $W$-plane whose two-sheets correspond to the two values of $\zeta$. In particular, the discrete transformation $\zeta \mapsto \zeta^{-1}$ moves between the two sheets over the same point in the $W$-plane.

When we come to apply these reductions to 6dCS theory, we expect to find a 4dCS theory on the correspondence space $M_{4} = \bCP^{1} \times \bR^{2}$. We have already seen that this may be parameterised by the coordinates $\{ \zeta , \xi , \bar{\xi} \}$, which transform under the discrete reduction as $\sigma^{\prime \prime} : (\zeta, \xi, \bar{\xi}) \mapsto (\zeta^{-1}, \xi, \bar{\xi})$. Combining this analysis with the $\bZ_{2}$-automorphism given earlier results in an equivariance condition on the 4dCS gauge field, 
\begin{equation}
    A(\zeta) = \eta \big( A(\zeta^{-1}) \big) \ . 
\end{equation}
In addition, the meromorphic 1-form of 4dCS theory is given by 
\begin{equation}
    \omega = \frac{1}{2} (X \wedge Y) \vee \Omega = \frac{1}{2} \, \frac{1 - \zeta^{2}}{\zeta^{2}} \dr \zeta = \dr W \ . 
\end{equation}
This is the appropriate 1-form to recover the 2dPCM and reproduces the setup described in \cite[\S 11]{Costello:2019tri}. We have therefore demonstrated how to recover the 2d symmetric-space PCM and 4dCS branch cut defects from 6dCS theory.